\documentclass[useAMS,usenatbib]{mn2e}
\usepackage{epsfig}
\usepackage{graphicx}
\usepackage{txfonts}
\usepackage{booktabs}

\def\vpsi{\bf \Psi }
\def\vq{\bf q}
\def\vx{\bf v}

\title[Cosmic voids detection without density measurements] {Cosmic voids detection without density measurements}

\author[A. Elyiv et al.]
{Andrii Elyiv$^{1,2}$\thanks{E-mail: andrii.elyiv@unibo.it}, Federico Marulli$^{1,3,4}$,  Giorgia Pollina$^{1,5}$, Marco Baldi$^{1,3,4}$, \newauthor Enzo Branchini$^{6,7,8}$, Andrea Cimatti$^{1}$ and Lauro Moscardini$^{1,3,4}$\\ \\ 
$^{1}$Dipartimento di Fisica e Astronomia, Universit\`a di Bologna, viale Berti Pichat 6/2, 40127  Bologna, Italy\\
$^{2}$Main Astronomical Observatory, Academy of Sciences of Ukraine, 27 Akademika Zabolotnoho St., 03680 Kyiv, Ukraine\\
$^{3}$INAF - Osservatorio Astronomico di Bologna, via Ranzani 1, 40127 Bologna, Italy\\
$^{4}$INFN - Sezione di Bologna, viale Berti Pichat 6/2, 40127 Bologna, Italy\\
$^{5}$University Observatory Munich, Ludwig-Maximilian University, Scheinerstr. 1, 81679, Munich, Germany\\
$^{6}$Dipartimento di Fisica, Università degli Studi Roma Tre, via della Vasca Navale 84, 00146 Rome, Italy\\ 
$^{7}$INAF - Osservatorio Astronomico di Roma, Monte Porzio Catone, Italy\\ 
$^{8}$INFN - Sezione di Roma Tre, via della Vasca Navale 84, 00146 Rome, Italy\\
}

\date{Released 2014 October 15}
\pagerange{\pageref{firstpage}--\pageref{lastpage}} \pubyear{2014}

\def\LaTeX{L\kern-.36em\raise.3ex\hbox{a}\kern-.15em
    T\kern-.1667em\lower.7ex\hbox{E}\kern-.125emX}

\begin{document}

\label{firstpage}

\maketitle

\begin{abstract}

{ Cosmic voids are effective cosmological probes to discriminate among
  competing world models.  Their
  identification is generally based on density or geometry criteria
  that, because of their very nature, are prone to shot noise. We propose two void finders that are based
  on dynamical criterion to select voids in Lagrangian coordinates and minimise the impact of sparse
  sampling. The first approach exploits the Zel'dovich approximation
  to trace back in time the orbits of galaxies located in voids
  and their surroundings, the second uses the observed
  galaxy-galaxy correlation function to relax the objects' spatial
  distribution to homogeneity and isotropy.  In both cases voids are
  defined as regions of the negative velocity divergence, that can be regarded as sinks of the back-in-time streamlines of the mass tracers.  
To assess the performance of our
  methods we used a dark matter halo mock catalogue {\small CoDECS}, and compared the results with those obtained with the {\small ZOBOV} void finder.  
We find that the void divergence profiles
  are less scattered than the density ones and, therefore, their stacking constitutes a more accurate cosmological probe. 
The significance of the divergence signal in the central part of voids obtained from both our finders is $60\%$ higher than 
for overdensity profiles in the {\small ZOBOV} case.  The ellipticity of the stacked void measured in the divergence field 
is closer to unity, as expected, than what is found when using halo positions.  Therefore our void finders are complementary 
to the existing methods, that should contribute to improve the accuracy of void-based cosmological tests.}

\end{abstract}


%%%%%%%%%%%%%%%%%%%%%%%%%%%%%%%%%%%%%%%%%%%%%%%%%%%%%%%%%%%%%%%%%%%%%%%%%%%%%%%%%%%

\section{Introduction}

Cosmic voids are regions of low density that occupy a significant
fraction of the volume in the Universe.  The first observational evidence
of giant voids was obtained more than 30 years ago
\citep{1978MNRAS.185.357, 1978ApJ.222.784}. However, their systematic
and quantitative study came later with the advent of the first large
spectroscopic surveys \citep{1988IAUS..130..105H, 1989Sci...246..897G,
  1995PASP..107..790B}.  Now it is well established that void sizes
span a wide range of scales, from minivoids with diameters of about
3-5 Mpc in the Local Universe \citep{2006ApJ...653..969T} to the
supervoids with diameters of about 200 Mpc \citep{1995A&A...301..329L,
  2014arXiv1406.3622S}.  Due to large volume filling factor,
characteristic shape, dynamics and low density, they constitute unique
laboratories for astrophysics and cosmology.

In the framework of high energy astrophysics, voids can be regarded as
{\em highways} for propagating particles like cosmic rays and
neutrinos \citep{2012ApJ...758..101S, 2013ApJ...770...54M,
  2014ApJ...789...84K}, in which the presence of non-zero
extragalactic magnetic fields or extragalactic light is still an open
issue \citep{2009PhRvD..80b3010E, 2012PhRvL.108q1301D,
  2013MNRAS.429L..60B, 2013arXiv1303.2121A}.  Voids are also important
for testing galaxy evolution models since they allow to study the
evolution of isolated objects and assess, by comparison, the influence
of the environment \citep{2001ApJ...557..495P, 2006MNRAS.369..335P,
  2009AN....330.1004V, 2010ASPC..421..107S,
  2014ApJ...780...88N}.  Indeed, the deficit of dwarf galaxies in
nearby voids \citep{2001ApJ...557..495P, 2011AJ....141..204K,
  2013AstBu..68....1E} with respect to the theoretical prediction,
often dubbed as the {\em Void Phenomenon}, is still regarded as a
potential evidence against the cold dark matter (CDM) scenario.

Finally, and most importantly, voids are very relevant for cosmology.
They have an impact on the Cosmic Microwave Background and on the weak
lensing signal. Anomalies like the Cold Spot could be explained  
as the result of the Integrated Sachs-Wolfe effect over large voids
\citep{1968Natur.217..511R, 2014arXiv1406.3622S, 2014arXiv1407.1470K}.
Next generation galaxy surveys like those that will be carried out by
the Large Synoptic Survey Telescope \citep{2002SPIE.4836..154K} and by
the Euclid satellite \citep{2011arXiv1110.3193L, 2013LRR....16....6A}
are expected to detect gravitational lensing from medium-sized voids
with which one can constrain the void's mass density profile without
using galaxies and assumptions on their bias
\citep{2013PhRvD..88b4049I, 2013ApJ...762L..20K, 2014MNRAS.440.2922M,
  2014arXiv1404.1834C}.

Indeed, voids can be used as effective cosmological probes. Their
physical properties depend on the nature of dark energy (DE) and on
the primordial density field from which they have evolved
\citep{2009PhRvD..80j3515O, 2011PhRvD..83b3521D, 2012MNRAS.426..440B,
  2013PhRvL.111x1103S, 2014MNRAS.438.1603G}.  In particular, it has
been realised that their shape is very sensitive to the equation of
state of the DE component and that spurious ellipticity could be used
to constrain the amount of dark matter (DM) \citep{2010MNRAS.403.1392L}.  Additional
constraints on DM can also be obtained by measuring the outflow
velocities from voids in the nearby Universe
\citep{2012ApJ...744...43C, 2013AJ....146...86T}.  Finally, voids'
expansion history and shape have been used to test modified gravity
models \citep{Li2011, 2013MNRAS.431..749C,2014arXiv:1410.0133,2014arXiv1410.1510C}. 

Perhaps their most common and effective cosmological application is
through the so-called Alcock-Paczynski (AP) test
\citep{1979Natur.281..358A}, in which one measures the size of the
stacked void along and across the line of sight and looks for a
possible mismatch that would arise from assuming an incorrect
cosmological model \citep{2012ApJ...761..187S, 2014MNRAS.438.3177S}.
The accuracy of the AP test has been discussed in comparison to that
of other cosmological tests by \citet{2012ApJ...754..109L}.  The first
application of the method by \citet{2012ApJ...761..187S} considered
voids extracted from the SDSS DR-7 galaxy survey
\citep{2009ApJS..182..543A} using a modified version of the ZOnes Bordering On Voidness ({\small
  ZOBOV}) algorithm \citep{2008MNRAS.386.2101N}. With $10$ stacked
voids detected out to $z = 0.36$ they failed to detect any significant
distortions. In a more recent attempt by \citet{2014arXiv1404.5618S}
that used the SDSS-DR10 LOWZ and CMASS galaxy samples
\citep{2014ApJS..211...17A}, the AP signal corresponding to a mass
density parameter $\Omega_{M}\simeq0.15$ has been detected.

The accuracy of the AP test depends on the ability to identify voids
in a reliable and robust way.  This is not an easy task, as
demonstrated by the fact that many different void finders have been
proposed over the years, many of which cross-compared in the
Aspen-Amsterdam contest \citep{2008MNRAS.387..933C}.
\cite{2010MNRAS.403.1392L} proposed to classify void finders in three
different categories depending on the type of criterion adopted for
the identification.  The first class is based on a density criterion
and defines voids as regions empty of galaxies or with local density
well below the mean \citep{2003MNRAS.344..715G, 2005MNRAS.360..216C,
  2007MNRAS.375..184B, 2013AstBu..68....1E}.  The second class uses
geometry criteria and identifies voids as geometrical structures like e.g.
spherical cells, polyedra, etc.  \citep{2002MNRAS.330..399P,
  2005MNRAS.360..216C, 2006MNRAS.367.1629S, 2007MNRAS.380..551P,
  2008MNRAS.386.2101N, 2014arXiv1410.0355L}. The third class is that of finders based on
dynamical criteria in which galaxies are considered as test particles
of the cosmic velocity field and not as tracers of the underlying mass
distribution \citep{2006MNRAS.375.489,2010MNRAS.403.1392L}.

Finders of the first two classes identify voids from galaxies'
Eulerian positions. This strategy has two disadvantages. First of all,
galaxies are used as mass tracers and some biasing prescription has to
be adopted to specify the relation between the galaxy and the mass
density field. This might play a significant role when attempting to
use cosmic voids to distinguish different cosmological scenarios, in
particular different theories of gravity or couplings in the dark
sector, as these models might feature a non-standard evolution of the
halo bias \citep[][]{2014Pollina}.  Second, voids are by definition
low density regions. Any method that uses galaxies to identify voids
is then prone to shot noise error. \cite{2014arXiv1407.1295N}
  showed that the naive strategy of measuring density in spherical
  shells around voids leads to a large number of empty shells and,
  consequently, a systematic underestimate of the density profile.
An additional drawback related to the Eulerian nature of these methods
is that they identify structures in a broad range of dynamical
regimes, which complicates the comparison with theoretical
predictions.

An advantage of void finders of the third class is that they can be
defined in Lagrangian coordinates, which greatly alleviates the shot
noise problem and eases the theoretical interpretation of the results.
The disadvantage, however, is that voids selected cannot be readily
compared with those obtained with the other methods. Perhaps, the best
worked out example of dynamical Lagrangian void finder is the {\small
  DIVA} method proposed by \citet{2010MNRAS.403.1392L}. This technique
uses the Monge-Amp\`{e}re-Kantorovitch method
\citep{2003MNRAS.346..501B} to reconstruct galaxy orbits back in time
starting from their Eulerian position, and identifies voids in
Lagrangian coordinates by looking at the regions where the divergence
of the displacement field is positive.  Assuming approximate
correspondence between the divergence field and linearly extrapolated
initial density field, they made predictions on void statistics and
local ellipticities defined from the curvature of the divergence
field. Dynamical void finders from the last class are very promising
for the probe of precision cosmology.

In this work we propose two 
%new 
dynamical void finders of the third class. Both methods aim at
reconstructing galaxies' Lagrangian positions by randomising the
Eulerian ones. The randomisation, however, is achieved in two
different ways. The first exploits galaxy clustering, and more
specifically the observed two-point correlation function, and relaxes
the system to a homogeneous and isotropic distribution using, in
reverse, an annealing scheme similar to the one proposed by
\cite{1997CIS.186.467} and commonly used in solid state physics to
create samples with specified spatial correlation properties. The
  second method is not completely new and actually quite similar to
  the {\small DIVA} method, except that it uses the Path Interchange
  Zel'dovich Approximation (PIZA) \citep{1997MNRAS.285..793C} to
  reconstruct galaxy orbits that, in this case, are simply straight
  lines. In this respect, more accurate reconstruction schemes could
be used, like those that minimise the action of a non-linear system
\citep{1989ApJ...344L..53P, 2000MNRAS.313..587N}. However, we decided
to adopt the PIZA scheme since it is fast, and easy to implement.

The randomisation procedure allows to build a displacement field which
approximates the cosmic velocity field and enables to trace the
buildup of cosmic structures, including voids. The latter are defined
as sinks in the time reverse streamline of galaxies. This is the
criterion that we use to locate voids in the Lagrangian coordinates.
We test our void finders using the catalogue of DM haloes extracted
from a high-resolution N-body simulation \citep{2012MNRAS.422.1028B} at $z=0$,
and compare our results with those obtained when we use the {\small
  ZOBOV} void finder \citep{2008MNRAS.386.2101N}. The quantitative
comparison focuses on voids' statistics, as well as on physical
properties such as their mass density profile and ellipticities.

The paper is organised as follows.  In \S\ref{sec:catalogue} we
describe the halo catalogues used in this work. In
\S\ref{sec:finders} we present the two proposed void finders based on
the two-point correlation function (\S\ref{sub:UVF}) and on the
Lagrangian Zel'dovich approximation (\S\ref{sub:LZ}).  The procedure
used for the final void identification is described in
\S\ref{sec:voidid}.  In \S\ref{sec:results} we analyse the properties
of the voids selected with the two methods, and compare them with the
ones detected by the {\small ZOBOV} finder. The conclusions are drawn
in \S\ref{sec:conclustions}.

%%%%%%%%%%%%%%%%%%%%%%%%%%%%%%%%%%%%%%%%%%%%%%%%%%%%%%%%%%%%%%%%%%%%%%%%%%%%%%%%%%%

\section{The DM halo catalogue}
\label{sec:catalogue}

In order to test our void finder algorithms, and to compare them with
those obtained with the public code {\small ZOBOV}, we use a halo
catalogue extracted from the {\small CoDECS} simulations \citep{2012MNRAS.422.1028B}, that are
the largest suite of publicly available\footnote{See www.marcobaldi.it/CoDECS} cosmological and hydrodynamical simulations
of interacting Dark Energy cosmologies \citep{Wetterich_1995, Amendola_2000} to date.
The CoDECS runs include five different models of DE interaction, plus a reference
  $\Lambda $CDM run.

In this work we focus on the standard $\Lambda $CDM run and plan to explore the
other interacting Dark Energy models included in the {\small CoDECS} Project in a future work aimed at assessing the impact of the dark sector interactions
on voids' properties
\citep{2014Pollina}. Specifically, we have considered the {\small
  H-CoDECS} run, a hydrodynamical N-body simulation with a comoving
box size of $80$ Mpc$/h$ and a total number of particles (gas + CDM)
of $N_{p}=2\times 512^3\approx 2.7\times 10^{8}$. The CDM mass
resolution is $m_{c} = 2.39\times 10^{8}$ M$_{\odot}/h$, and the
gravitational softening was set to $\epsilon _{g} = 3.5$ kpc$/h$,
corresponding to about $1/40th$ of the mean inter-particle separation,
$\bar{d}$.  CDM haloes have been identified by means of a
Friends-of-Friends (FoF) algorithm \citep{Davis_1985} with linking
length $\lambda = 0.2\times \bar{d}$. This procedure has been applied
to the particle distribution by linking the CDM particles as primary
tracers of the local mass density, and then attaching baryonic
particles to the FoF group of their nearest neighbour.  
The {\small CoDECS} public repository provides also spherical overdensity 
catalogs computed with the {\small SubFind} algorithm \citep[][]{2001MNRAS.328..726S}, 
but for the present analysis we will consider only the FoF catalogs.
In this work
we focus on the $z=0$ output. At this epoch the halo catalogue
consists of $116129$ haloes with a minimum mass of $7.657\times
10^{10}{\rm M_{\odot}}/h$, corresponding to an average halo density of
$\rho_{\rm h}=0.23~({\rm Mpc/h})^{-3}$ and to a corresponding mean
inter-halo separation of $1.64~{\rm Mpc}/h$.  The average halo
density is high enough to identify small voids with sizes above $3 {\rm
  Mpc}/h$, while the large volume of the sample, $5.12\times 10^{5}
{\rm Mpc}^{3}/h^{3}$, may provide a wide range of void sizes up to
$\sim 20~{\rm Mpc}/h$.

%%%%%%%%%%%%%%%%%%%%%%%%%%%%%%%%%%%%%%%%%%%%%%%%%%%%%%%%%%%%%%%%%%%%%%%%%%%%%%%%%%%

\section{
%New void finders
Dynamical void finders}
\label{sec:finders}

The basic idea behind the two dynamical void finders that we present
in this Section is rather simple.  Let us consider a volume of the
Universe characterised by large scale structures in the DM component
and probed by ``particles'' like haloes or galaxies. Here we propose
to use these as test particles, not mass tracers, and trace their
orbits back in time to a homogeneous and isotropic initial
distribution, that is to reconstruct their Lagrangian positions $\vq$.
This is done in two ways: {\em i)} by exploiting the different correlation
properties of the initial and final particle distributions and {\em ii)} by
assuming some dynamical approximation that allows to solve the mixed
boundary problem of a system of particles with known final positions
and well defined initial statistical properties.  The practical
implementation of these ideas are described in \S\ref{sub:UVF} and
\S\ref{sub:LZ}, respectively.

After reconstructing Lagrangian positions, $\vq$, we obtain the
particle displacement field, $\vpsi$, by simply connecting them to the
Eulerian positions, $\vx$, hence assuming straight orbits:
$\vq-\vx=\vpsi(\vq)$. The divergence of the displacement field,
$\Theta \equiv \nabla \cdot \vpsi$, is associated to the mass density
field and used to identify voids as sinks of mass streamlines in
time reverse variables.  We note that this approach is similar to that
of \cite{2013MNRAS.434.1192R}, in which a positive divergence in the
galaxy flow was one of the conditions used to identify voids.

\begin{figure*}
\begin{tabular}{c c}
\hspace*{-3.2cm}\epsfig{file=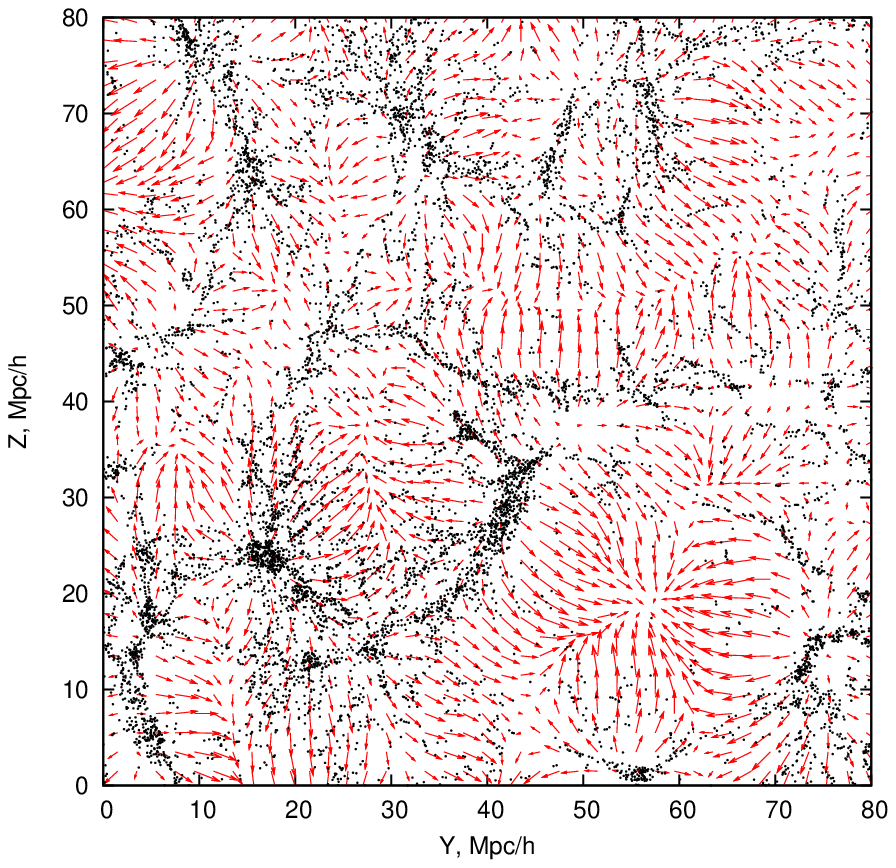,width=12.1cm}&\hspace*{-4.6cm}\epsfig{file=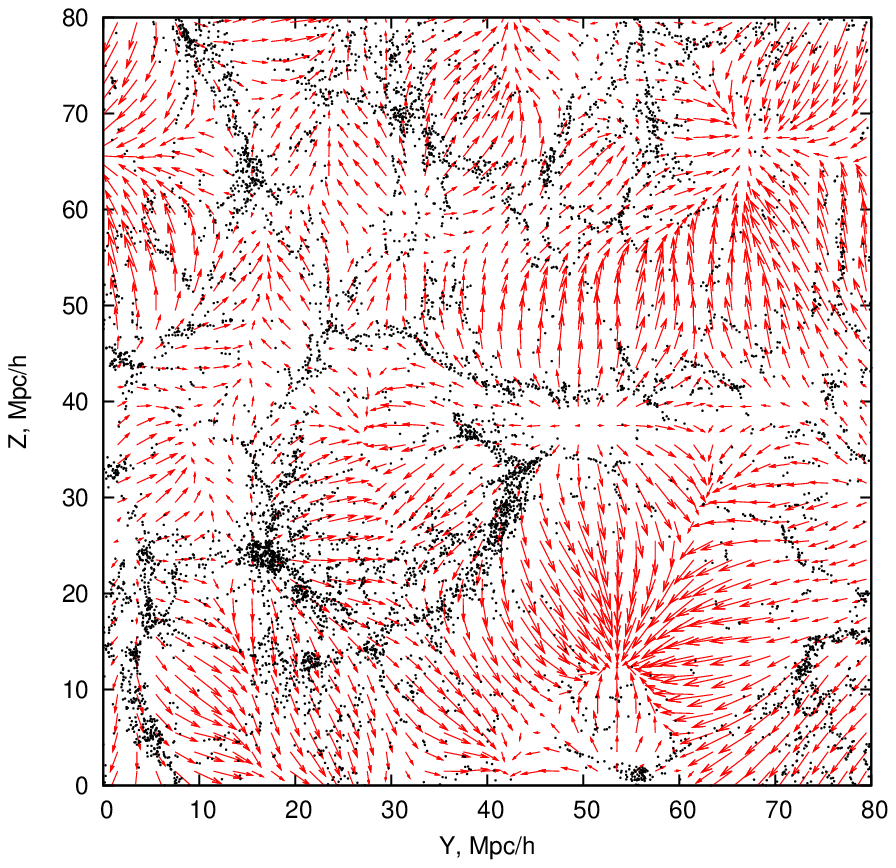,width=12.1cm}\\
\hspace*{-0.9cm}\epsfig{file=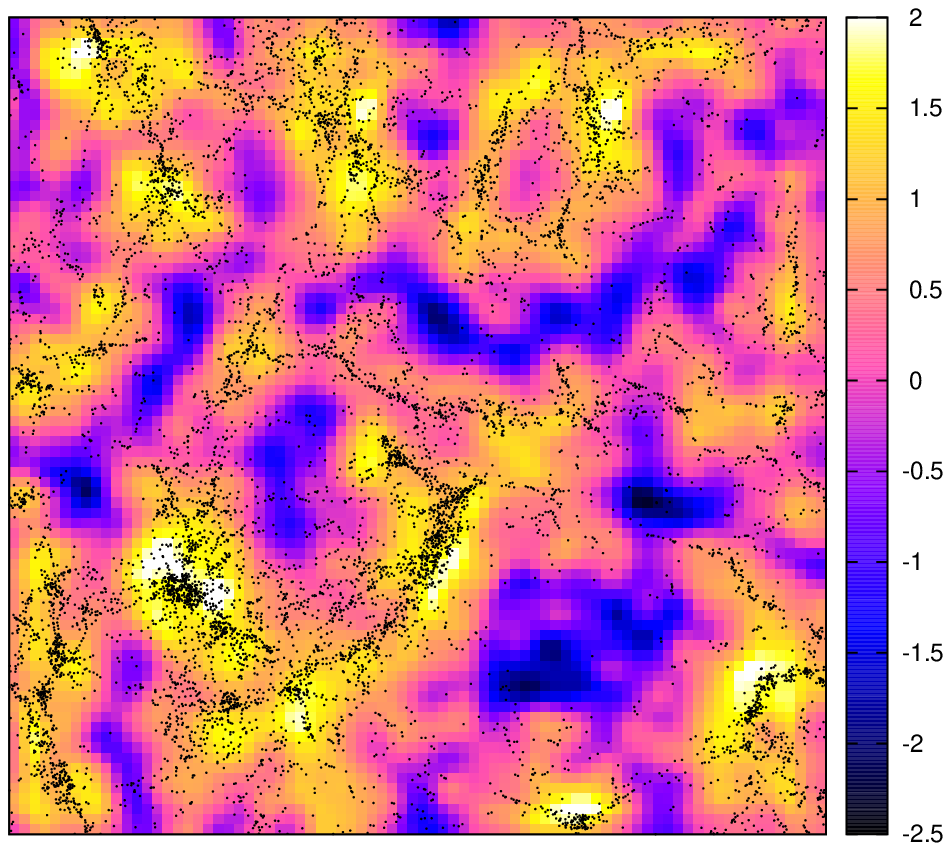,width=11.5cm}&\hspace*{-2.3cm}\epsfig{file=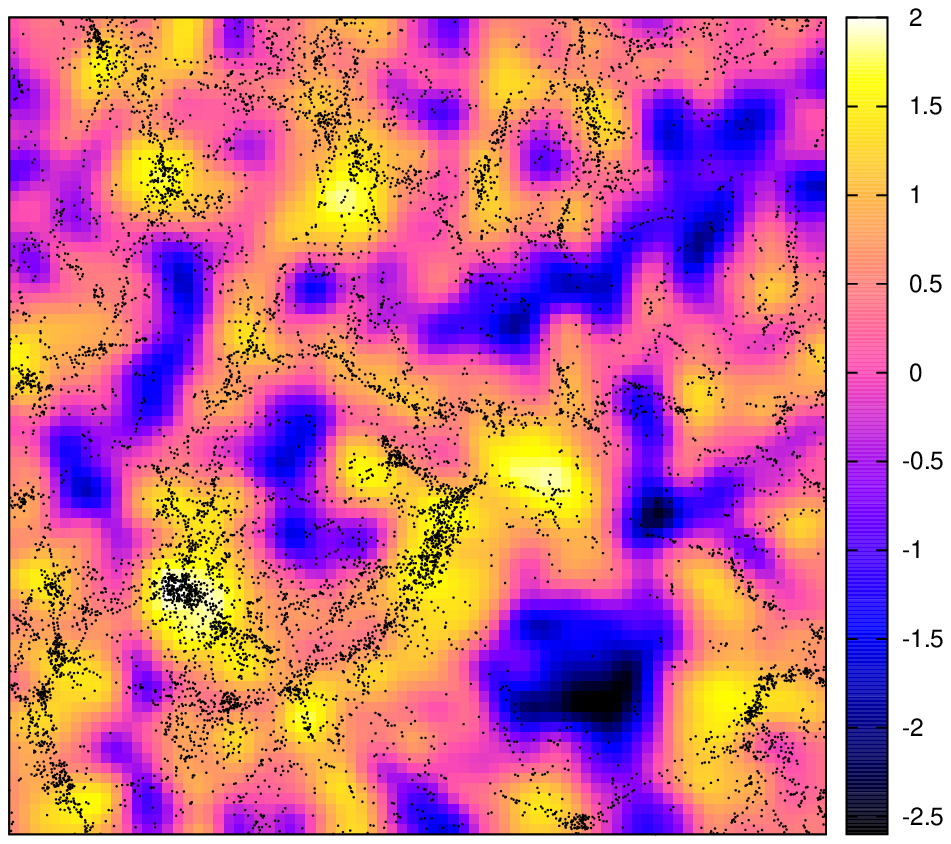,width=11.5cm}\\
\end{tabular}
\caption{The reconstructed displacement field ${\bf \Psi}$ ({\em top
    panels}) and its divergence $\Theta$ ({\em bottom panels}),
  obtained with the two void finders proposed in this work, the UVF
  ({\em left-hand panels}) and the LZVF ({\em right-hand panels}).
  The size of the displayed region is $80{\rm x}80~{\rm Mpc}/h$, with
  a thickness of $5~{\rm Mpc}/h$. Black dots represent dark matter
  haloes. The amplitude of the vector field components (red arrows) is
  reduced by a factor of $0.75$, for visual clarity.}
\label{fig:map}
\end{figure*}

%%%%%%%%%%%%%%%%%%%%%%%%%%%%%%%%%%%%%%%%%%%%%%%%%%%%%%%%%%%%%%%%%%%%%%%%%%%%%%%%%%%

\subsection{The Uncorrelating Void Finder}
\label{sub:UVF}

At the present epoch the spatial distribution of galaxies is highly
inhomogeneous. Deviations from the homogeneity are conveniently and
readily characterised by their spatial two-point correlation function
that measures the departure from a purely Poisson distribution.  On
the other hand, at early epochs the distribution of matter and of any
test particles like galaxies is supposed to be highly homogeneous,
with no spatial correlation.  This suggests that a practical way to
trace galaxy orbits back in time, at least in a statistical sense,
could be that of relaxing their present spatial distribution to
homogeneity, practically defined as a state in which the correlation
function at all separations is zero.  An effective way to achieve this
is by applying the annealing method of \cite{1997CIS.186.467} in
reverse, effectively moving the system away from its minimum energy
configuration.  We note that this method has been used to generate
mock galaxy catalogues with a specified two-point correlation function
\citep{2011ApJ...727...45S}.  In our iterative algorithm we allow for
a small random displacement of one halo along a random direction with
the length $0.5~{\rm Mpc}/h$.  If the amplitude of the two-point
correlation function decreases then we place the halo at the new
position. Otherwise we keep it at the original position. Then we move
to the next halo and repeat the procedure for all haloes in the
catalogue.  Then we move to the next iteration and start again with
the first halo.  Iterations stop when the amplitude of the two-point
correlation function is consistent with zero within the uncertainties
at all separations.  More quantitatively, iterations stop when
$\sum_{i=1}^{N} \xi(r_{i})^{2}/N_{b}<\varepsilon_{lim}$, where
$\xi(r_{i})$ is the amplitude of the two-point correlation function at
separation $r_{i}$, and $N_{b}$ is the total number of bins. In this
work we used $N=50$ logarithmic bins over the range of scales
$0.5<r[{\rm Mpc}/h]<50$ .  We set the tolerance parameter empirically
to the value $\varepsilon_{lim}=10^{-6}$, as a compromise between
reconstruction accuracy and computational cost.  For a fast
measurement of the two-point correlation function, we use a
linked-list based algorithm \citep{2012MNRAS.426.2566M,
  2013A&A...557A..17M}, specifically modified to update at each
iteration step only the number of pairs associated to the halo that is
moved in the procedure\footnote{Using $\sim116000$ haloes as test
  particles, the code requires $\sim100$ hours of CPU time to detect
  the voids, with an Intel Core i5, 520M@2.4 GHz CPU.}.

At the various steps of the iteration the haloes describe a random
walk from high density regions to low density ones. We compute the
displacement vector $\vpsi$ that connects the Eulerian and the
reconstructed Lagrangian positions at the end of the iteration. Since
the reconstruction is not unique, the Lagrangian positions, and
consequently the displacement field, depend on the random seed. We
perform $10$ reconstructions using different seeds and obtain the
average displacement field at the given position ${\bf q}$ by
Gaussian-weighting the individual displacement vectors at the
different reconstructed Lagrangian positions, that is 
\begin{equation}
\langle
\Psi({\bf q}) \rangle =\sum_{j}{\Psi_{j}\cdot
  \exp(-d_{j}^{2}/2 \sigma^{2})}/{\sum_{j} \exp(-d_{j}^{2}/2
  \sigma^{2})}.
\end{equation}
Here the 
%subscript $i$ indicates the generic Cartesian component of the average displacement vector, the 
subscript $j$ identifies the reconstructed Lagrangian positions in any
of the $10$ reconstructions and the sum runs over all haloes'
Lagrangian positions in the $10$ reconstructions, $d_{j}$ is the
distance of the halo $j$ to the given position ${\bf {q}}$. Such
a Gaussian smoothing is a needed procedure to convert discrete
displacement vectors which are computed at the Lagrangian positions of
galaxies to a contiguous vector field. The most direct way is Gaussian
smoothing with kernel scale
%comparable to the mean galaxy distances in the sample.
 %We choose a smoothing length of 
$\sigma=1.5~{\rm Mpc}/h$, that corresponds to the mean inter-halo
distance.  Finally, the displacement field is used to build the
divergence field $\Theta=\nabla \cdot \vpsi$ at the generic position
$(x,y,z)$ which, for convenience, we place at the nodes of a cubic
grid of size $1~{\rm Mpc}/h$.  We shall call this method the {\em
  Uncorrelating Void Finder} (UVF).

Fig.~\ref{fig:map} shows a slice of thickness $5~{\rm Mpc}/h$ across
the computational cube. The top left panel shows the Eulerian
positions of the haloes (black dots) superimposed to the projected
displacement field $\langle \vpsi \rangle$ (red arrows).  In the
bottom panel the same black dots are superimposed to the divergence
field. The colour code is set according to the amplitude of $\Theta$,
as indicated by the colour bar.

%%%%%%%%%%%%%%%%%%%%%%%%%%%%%%%%%%%%%%%%%%%%%%%%%%%%%%%%%%%%%%%%%%%%%%%%%%%%%%%%%%%

\subsection{The Lagrangian Zel'dovich Void Finder} 
\label{sub:LZ}

The second void finder that we consider here is based on the
Zel'dovich approximation \citep{1970A&A.....5...84Z} to the growth of
density fluctuations.  More specifically, we implement the PIZA method
of \citet{1997MNRAS.285..793C} to trace the orbit of the objects in a
self gravitating system by minimising its action, under the Zel'dovich
approximation.  Since objects have straight orbits, we simply connect
their Eulerian positions to those of a randomly distributed sample.
In each iteration we modify the connection, hence setting a new path
to a different grid point, and accept it if the total path, obtained
by summing the square of the individual paths, decreases. Since the
total path is proportional to the action, this system is relaxed to
the correct dynamical configuration. We will refer to this finder as
the {\em Lagrangian Zel'dovich Void Finder} (LZVF).

Again, the reconstructed positions depend on the choice of the random
seed used in the iterative procedure. 
%Therefore, in analogy with the
%first estimator, we obtain the solution by Gaussian-weighting the
%displacement fields obtained from $10$ different reconstructions.
The displacement fields obtained from $10$ different reconstructions were smoothed by a
Gaussian kernel of  radius 1.5 Mpc/h, i.e. the same one used in the UVF finder.
%Finally, the divergence field is obtained from the averaged
%displacement field.

The result is shown in the right panels of Fig.~\ref{fig:map}, that
are analogous to their counterpart on the left. Clearly both the
divergence and the displacement fields obtained with the two methods
are qualitative similar. The main features (coherent flows, regions of
large positive/negative divergence) are seen on both
reconstructions. However, their amplitude and spatial location do not
always coincide. We shall see in the next Sections what are the
impacts of these differences on the properties of the voids.

%%%%%%%%%%%%%%%%%%%%%%%%%%%%%%%%%%%%%%%%%%%%%%%%%%%%%%%%%%%%%%%%%%%%%%%%%%%%%%%%%%%

\section{Void Identification and Characterisation}
\label{sec:voidid}

By definition the divergence of the vector field represents the
density, $\Theta$, of sources or sinks of that field at a given point:
\begin{equation}
\Theta = {\bigtriangledown\vpsi}=\sum_{i=1}^{3}\frac{\partial
  \Psi_{i}}{\partial x_{i}} \, ,
\end{equation}
where $x_{i}$ are the Cartesian coordinates. We calculate the
divergence numerically in a grid with step $1~{\rm Mpc}/h$. The colour
maps in the lower panels of Fig.~\ref{fig:map} show the divergence
field assessed with the two proposed void finders, UVF (left) and LZVF
(right). Clearly, there is a strong correlation between the
distribution of the Eulerian halo positions and the magnitude of
divergence calculated in Lagrangian coordinates.

\begin{figure}
\hspace*{-0.3cm}\includegraphics[scale=0.7]{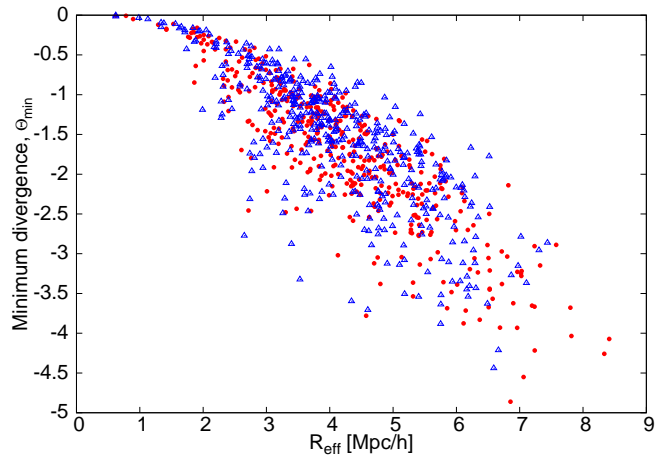}
\caption{Relation between the effective subvoid radii, $R_{eff}$, and
  the divergence minima, $\Theta_{min}$, for the two void finders, UVF (blue triangles)
  and LZVF (red circles).}
\label{fig:thetaR}
\end{figure}

We use the dimensionless quantity $\Theta$ to characterise the voids,
instead of the local density. Specifically, the local minima of the
divergence field are used to localize the voids. In our approach, a
local minimum is a grid node where $\Theta$ is lower than in all the
adjacent $3^{3}-1$ nodes around it. Moreover, we consider only the
local minima with negative divergence.  Each local minimum identifies
a subvoid. To determine its shape we use the well known watershed
technique \citep{2007MNRAS.380..551P}.  We fill the basin starting
from the local minimum. When the ``water'' level of a subvoid reaches
the ones of ajacent subvoids, we do not join these regions but just
put a wall between them and continue by filling until all grid nodes
with $\Theta<0$ are not involved.  In such a way the divergence field
divides the space between subvoids around each local minimum.  We do
not use any other additional parameters or criteria for the void
identification, as the average or minimum halo density.  This approach
provides a strict definition of subvoids, which can be considered as
the sinks of the back-in-time streamlines of the mass tracers. The
subvoids are the roots in the tree hierarchy, and have no parent
voids. We do not consider subvoids that lie close to the boundary of
the sample, as their volumes would be underestimated.

To assess the subvoid as well as void sizes, we estimate the spherical equivalent radius
$R_{eff}=\left(\frac{3V_{void}}{4\pi}\right)^{1/3}$, where $V_{void}$
is the volume of the void. We find a moderate correlation between the
values of minimum divergence, $\Theta_{min}$, and the effective radii
of the corresponding subvoids, as shown in Fig.~\ref{fig:thetaR}. This
is expected: the smaller is the minimum divergence, the larger is the
subvoid volume around.  It can be interpreted according to the large
scale structure growing scenario. The larger is a galaxy escaping
velocity at a given place, the more effective is the void expanding
there, as confirmed theoretically for isolated perturbations of the
density field \citep{1986ApJ...304...15B, 2004MNRAS.350..517S}. This
result is in agreement with the findings by
\citet{2014arXiv1407.1295N}, obtained using the {\small ZOBOV}
watershed transform algorithm.

 The construction of the void hierarchical tree is not
  unique. Here we have adopted a procedure based on merging adjacent
  voids and set the minimum effective radius of accepted voids,
  $R_{lim}$. Voids are said to be adjacent when they have been
  assigned two nearby grid points. The procedure is the following:\\ -
  Voids are sorted by their effective radius from the largest to the
  smallest.\\ - Voids with an effective radius larger than $R_{lim}$
  are kept.\\ - Voids with radius below $R_{lim}$ that are isolated
  (that is with no adjacent void) are discarded.\\ - Voids with radius
  below $R_{lim}$ with one or more adjacent voids, are merged to the
  adjacent one with the largest radius.  The effective radius of the
  new merged void is calculated from the sum of the void
  volumes that are included in the merged void.  We repeat this
  procedure until all adjacent voids have radii below
  $R_{lim}$.\\ - After the merging procedure we keep all objects with
  radius larger than $R_{lim}$.\\

%A large void can be formed by many smaller subvoids.  To go through
%the void hierarchical tree, we set the minimum effective radius of
%accepted voids, $R_{lim}$.  If the radius of a void is smaller than
%$R_{lim}$, we join it with an adjacent one with a larger radius.  We
%repeat this procedure until all voids have $R_{eff}>R_{lim}$.  The
%rest of isolated voids with $R_{eff}<R_{lim}$ are removed.
 
A crucial issue to be addressed when reconstructing void profiles and
stacking is how to define the void centres.  We consider four
alternative definitions. The first one is the geometrical centre
defined as follows:
\begin{equation}
{\bf r_{G}}=\frac{\sum^{n}_{i=1}{r_{i}}}{n} \, ,
\end{equation}
where $r_{i}$ are the coordinates of all $n$ grid nodes that belong to
the void, that is ${\bf r_{G}}$ is the centre of mass if the void
density is uniform.

The second definition is the barycentre of the haloes hosted in the
void:
\begin{equation}
{\bf r_{B}}=\frac{\sum^{N}_{j=1}{v_{j}}}{N} \, , 
\end{equation}
where $v_{j}$ are the coordinates of the $N$ haloes in the
void.

The third void centre is defined as the weighted centre over the
divergence field inside a given void:
\begin{equation}
{\bf r_{W}} = \frac{\sum^{n}_{i=1}\Theta(r_{i}) r_{i}}{\sum^{n}_{i=1}
  \Theta_{i}(r_{i})} \, ,
\end{equation}
where ${\bf r_{i}}$ are the coordinates of all $n$ grid nodes that
belong to the void.

The last definition considered in this work is the position of the
minimum of the divergence field inside a given void, ${\bf r_{M}}$.
In the following figures, we will refer to these four different
definitions of void centres as {\bf G}, {\bf B}, {\bf W}, {\bf M},
respectively.

%%%%%%%%%%%%%%%%%%%%%%%%%%%%%%%%%%%%%%%%%%%%%%%%%%%%%%%%%%%%%%%%%%%%%%%%%%%%%%%%%%%

\section{Results}
\label{sec:results}

Applying our two finders, UVF (\S\ref{sub:UVF}) and LZVF
(\S\ref{sub:LZ}), to the $z=0$ DM halo sample described in
\S\ref{sec:catalogue}, we construct a set of void catalogues. In the
following Sections we present the main properties of the selected
voids, and compare them to the ones detected by the {\small ZOBOV}
finder in \S\ref{sub:zobov}.

%%%%%%%%%%%%%%%%%%%%%%%%%%%%%%%%%%%%%%%%%%%%%%%%%%%%%%%%%%%%%%%%%%%%%%%%%%%%%%%%%%%

\subsection{Void statistics}

Using the two void finders, UVF and LZVF, we find in total $427$ and
$433$ subvoids, respectively, with effective radii up to $\sim 8.4
{\rm Mpc}/h$. We construct different void samples by cutting at
different $R_{lim}$. The main properties of the selected void
catalogues are summarized in Table \ref{table:1}. 

\begin{figure}
\hspace*{-0.9cm}\includegraphics[scale=0.38]{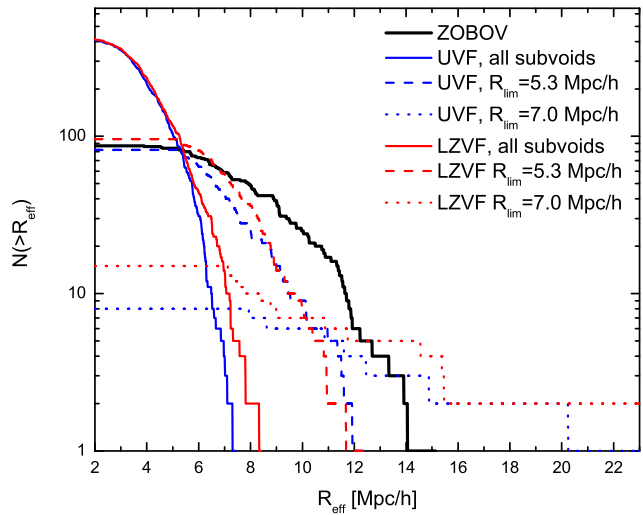}
\caption{The cumulative distribution of selected voids as a function
  of $R_{eff}$, for different values of $R_{lim}$,
  as indicated by the labels. Blue, red and black lines show the
  results obtained with UVF, LZVF and {\small ZOBOV}, respectively.}
\label{fig:distr}
\end{figure}

Fig.~\ref{fig:distr} compares the cumulative distributions of voids
selected with different finders, as a function of the effective radius
$R_{eff}$, and for three different values of $R_{lim}$. As expected,
these distributions become steeper as $R_{lim}$ decreases. The results
obtained with the two finders UVF and LZVF appear in reasonable
agreement. 

The use of a grid and of a smoothing of the displacement field could
in principle introduce a numerical bias in our results. We investigate
the impact of these issues by changing the grid steps and the
smoothing scale. An increasing of the grid step from $1$ to $2$ Mpc
has the effect of enhancing the median subvoid radii by $\sim 10\%$,
while the size of the larger merged voids and the divergence profiles
are almost unaffected. When increasing the smoothing scale from $1.5$
to $3 {\rm Mpc}/h$, the median radii of subvoids are enhanced by $\sim
30(70)\%$ for $R_{lim}=4(0)~{\rm Mpc}/h$, while the sizes of larger
voids are unchanged.

\begin{table}
\caption{Main properties of the void samples selected with UVF, LZVF
  (for different values of $R_{lim}$) and {\small ZOBOV}.}
\label{table:1}
\begin{tabular}{cccc}
\hline
void sample,                & number  & $R_{eff}$ range & median $R_{eff}$  \\
$R_{lim}$ [Mpc/h] &    of voids                  & [Mpc/h] & [Mpc/h] \\
\hline
UVF finder\\
all subvoids & 427  & 0.6 - 7.4 & 4.0 \\
2 & 405 & 2.0 - 7.4 & 4.1 \\ 
3 &  344 & 3.0 - 7.4  & 4.4 \\
4 &  211 & 4.0 - 8.0  & 5.4\\
5   & 103  & 5.0 - 11.6 & 6.8 \\
5.3 &  82  & 5.3 - 12.6 & 7.0         \\  
6 &  34 & 6.0 - 17.1  & 8.3 \\
7 & 8  & 7.0 - 28.7  & 12.0 \\
8 & 7 & 8.0 - 28.7 & 12.5 \\
\hline
LZVF finder \\
all subvoids &433     & 0.6-8.4 & 4.1  \\
2            & 412    & 2.0-8.4 & 4.2 \\
3 & 358    &3.0-8.5  & 4.4\\
4 & 233   &4.0-8.5  & 5.3\\
5   &  117   &  5.0-12.2 & 6.9 \\
5.3 &  96     & 5.3-12.6 &  7.3      \\  
6 & 46&    6.0-17.9 & 8.5\\
7 & 15      & 7.0-26.5  & 9.0\\
8& 5 &   8.0 - 32.2 & 17.5 \\

\hline
{\small ZOBOV} & 87 & 3.9 - 15.1& 8.2
\end{tabular}
\end{table}

\begin{figure*}
\begin{tabular}{c c}
\epsfig{file=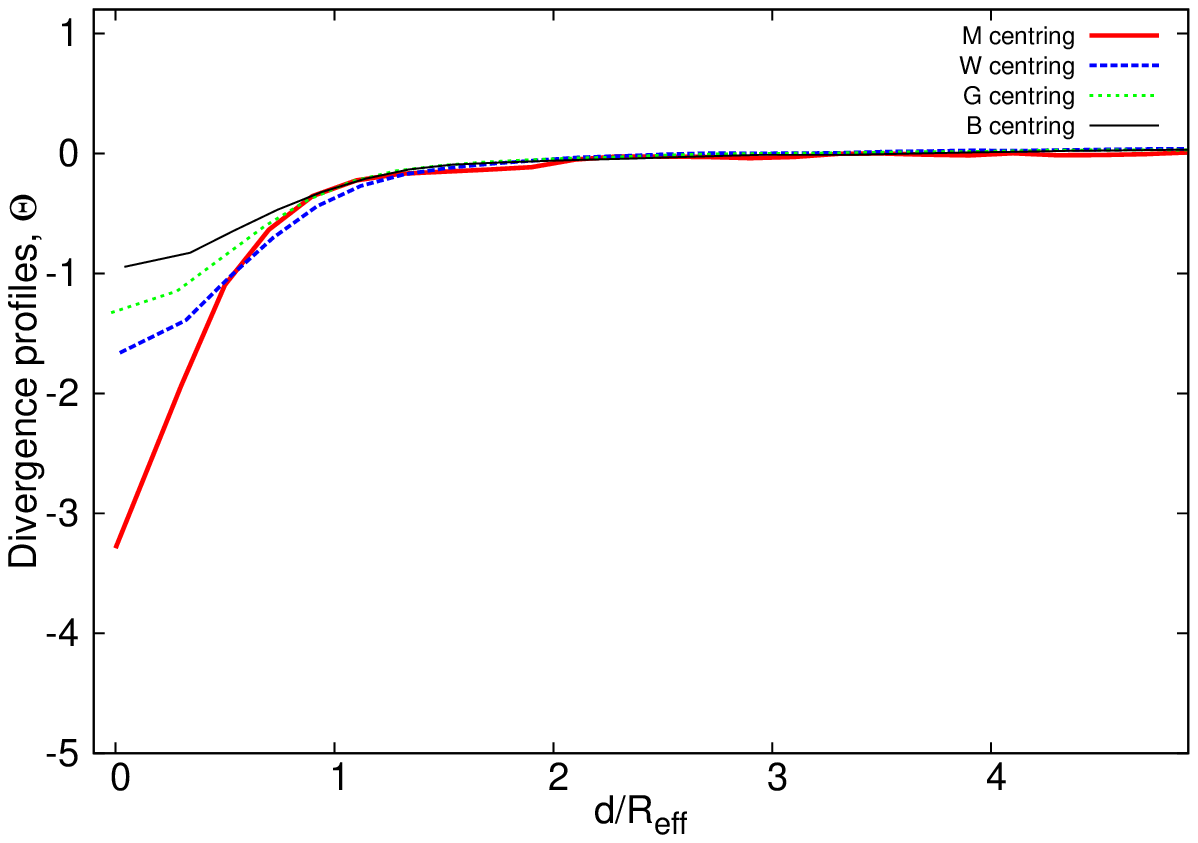,width=8.6cm}&\epsfig{file=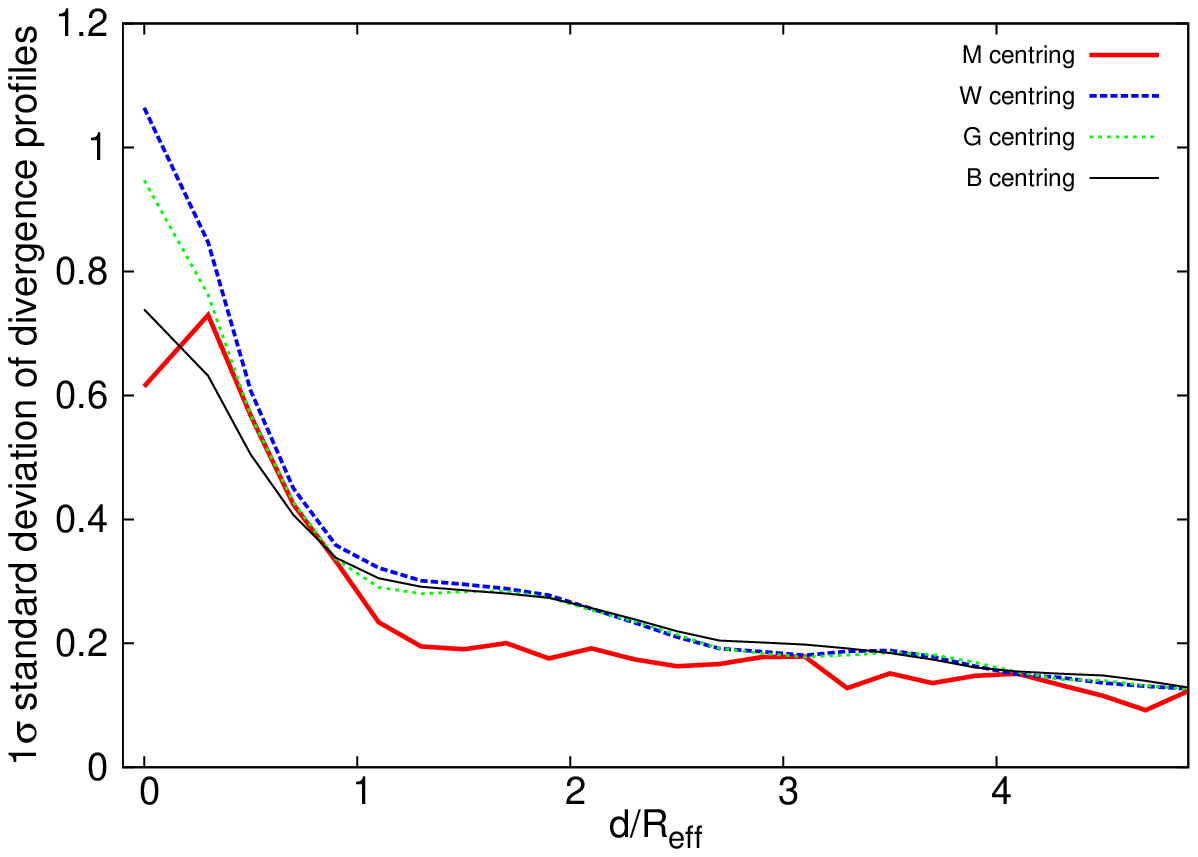,width=8.6cm}
\end{tabular}                                                             
\caption{Mean divergence profiles ({\em left-hand panel}) and
  corresponding $1\sigma$ deviations ({\em right-hand panel}) of the
  voids detected by the LZVF, as a function of the normalized radial
  distance, $d/R_{eff}$, and for different centrings, as indicated by
  the labels.}
\label{fig:divR}
\end{figure*}

\begin{figure*}
\begin{tabular}{c c}
\epsfig{file=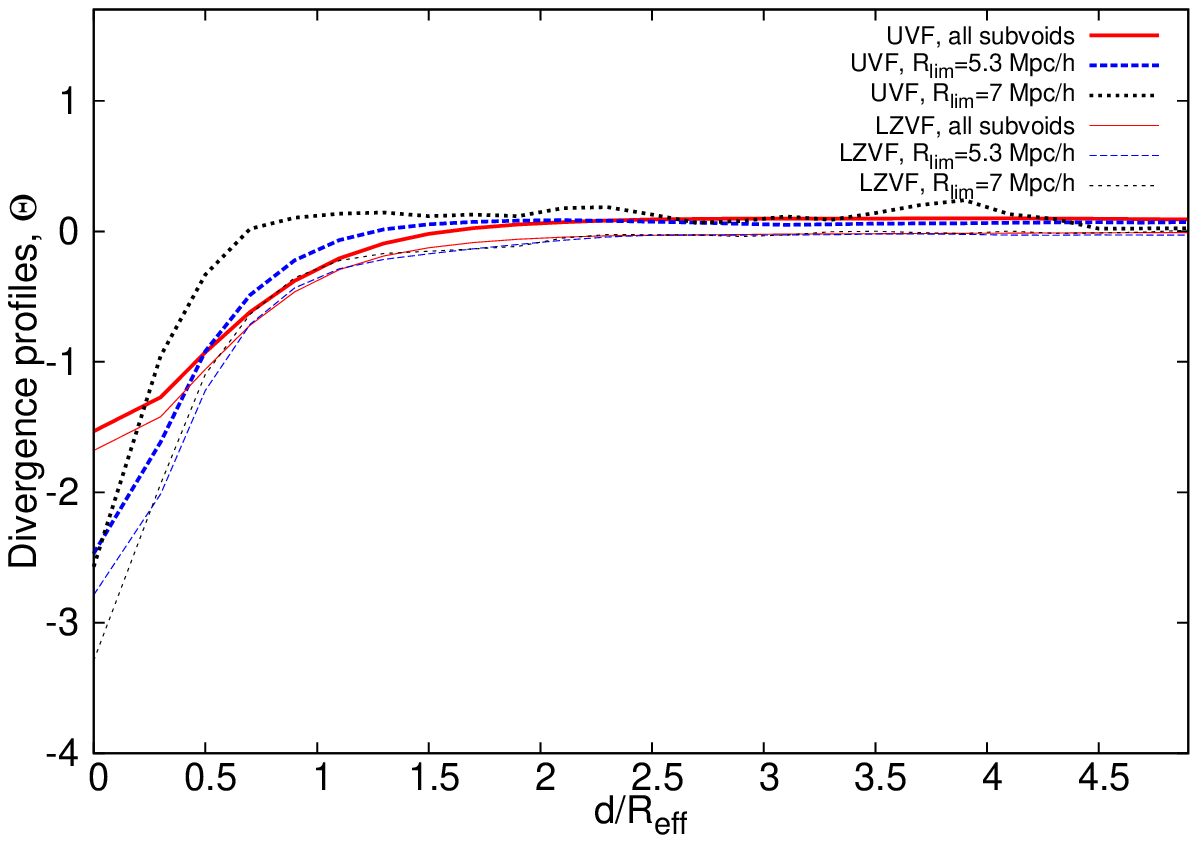,width=8.6cm}&\epsfig{file=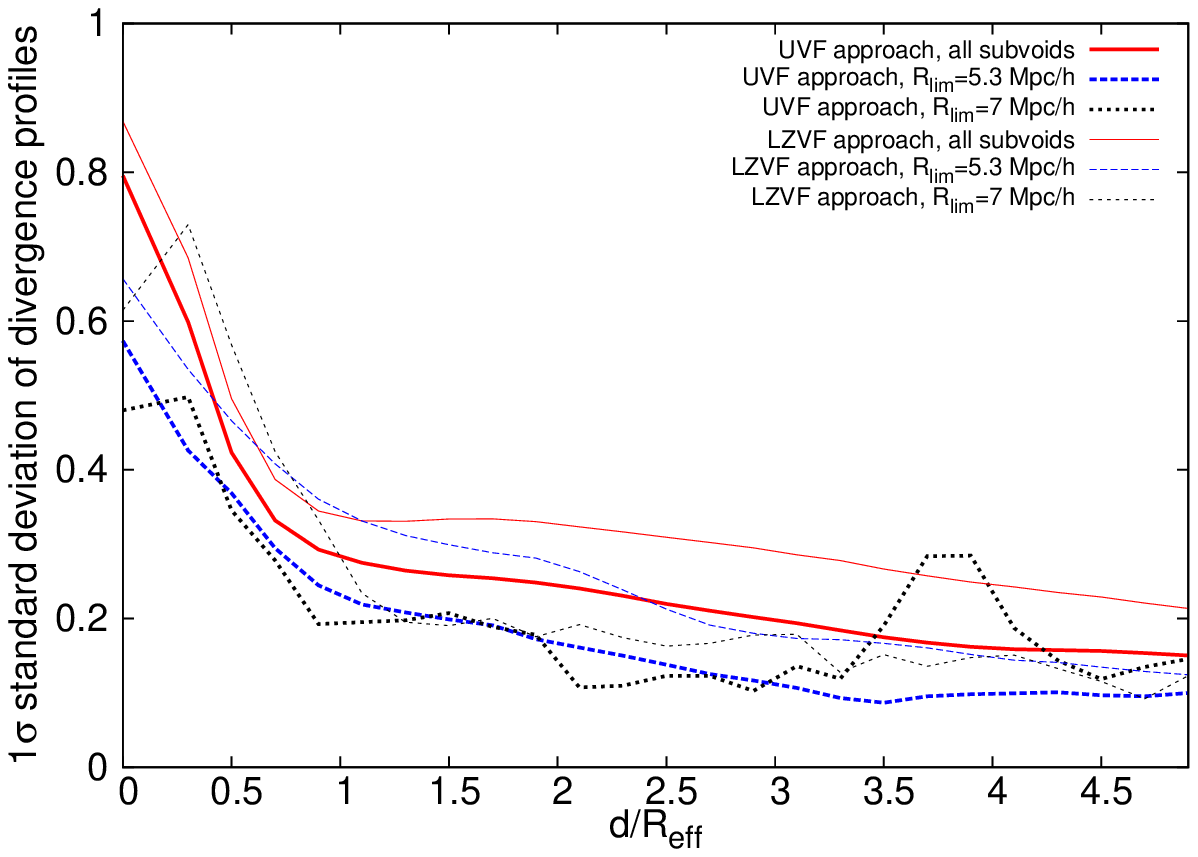,width=8.6cm}
\end{tabular}
\caption{Mean divergence profiles ({\em left-hand panel}) and
  corresponding $1\sigma$ deviations ({\em right-hand panel}) of the
  voids detected by the two finders UVF and LZVF, as a function of the
  normalized radial distance, $d/R_{eff}$, and for three different
  values of $R_{lim}$ ({\bf M} centring), as indicated by the
  labels.}
\label{fig:divV}
\end{figure*}

%%%%%%%%%%%%%%%%%%%%%%%%%%%%%%%%%%%%%%%%%%%%%%%%%%%%%%%%%%%%%%%%%%%%%%%%%%%%%%%%%%%

\subsection{Divergence profiles}

Fig.~\ref{fig:divR} shows the mean divergence profiles ({\em left
  panel}) and associated $1\sigma$ deviations ({\em right panel}) as a
function of the normalized distance $d/R_{eff}$ from the void
centres. In particular, this figure shows the impact of the different
definitions of void centres considered in this work. As it can be seen in
the left panel, the divergence profiles obtained with the {\bf M}
centring, that is by defining the void centres as the minima of the
divergence field, are the steepest in the central part and the least
noisy. Therefore, in the following analysis we will consider only this
definition.

Fig.~\ref{fig:divV} compares the results obtained with UVF and LZVF,
for three different values of $R_{lim}=0, 5.3$ and $7 {\rm Mpc}/h$ (with
{\bf M} centring). While the mean divergence profiles of the two void
finders are in close agreement, we note that the ones provided by UVF
are the least noisy, independent of $R_{lim}$.  As the selected voids
are not exactly spherical, the outer parts of these profiles can
include also regions outside the voids. The effect is however low, and
does not significantly impact our results.

%%%%%%%%%%%%%%%%%%%%%%%%%%%%%%%%%%%%%%%%%%%%%%%%%%%%%%%%%%%%%%%%%%%%%%%%%%%%%%%%%%%

\subsection{Overdensity profiles}

\begin{figure*}
\begin{tabular}{c c}
\epsfig{file=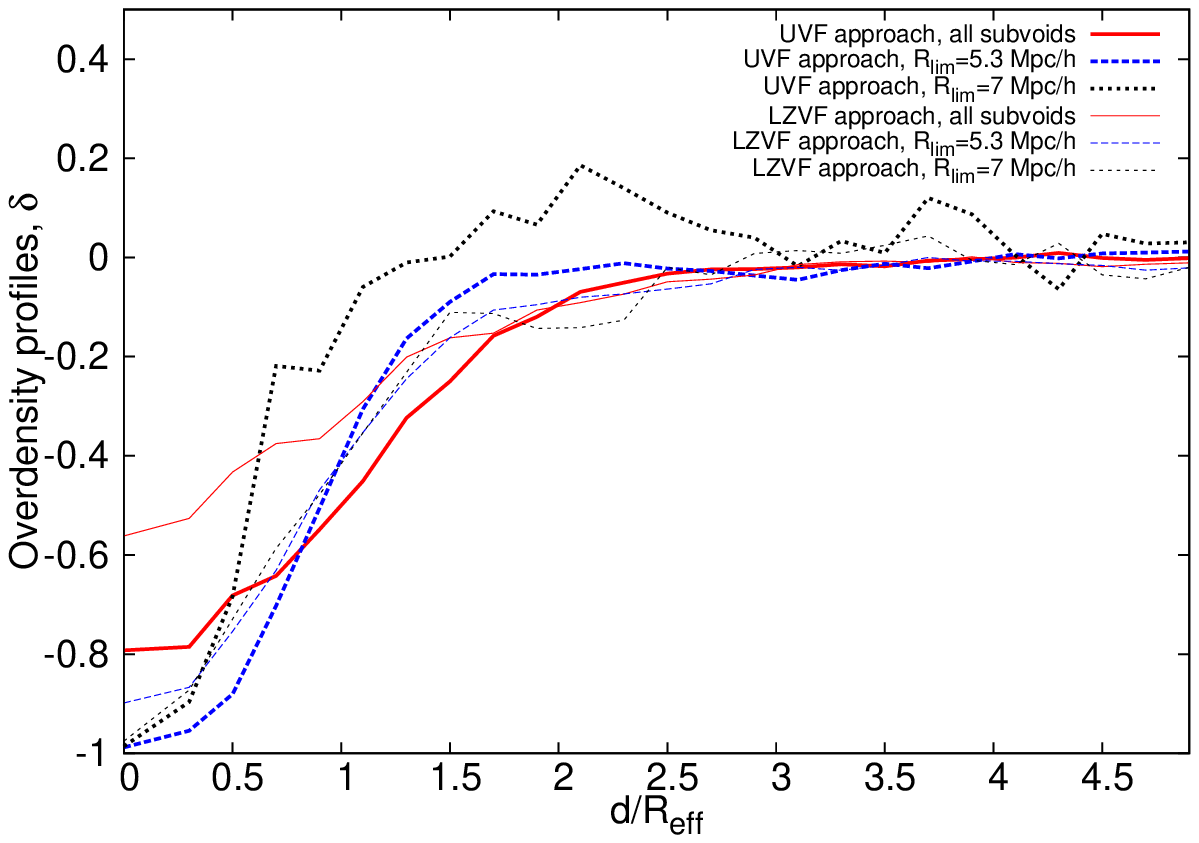,width=8.6cm}&\epsfig{file=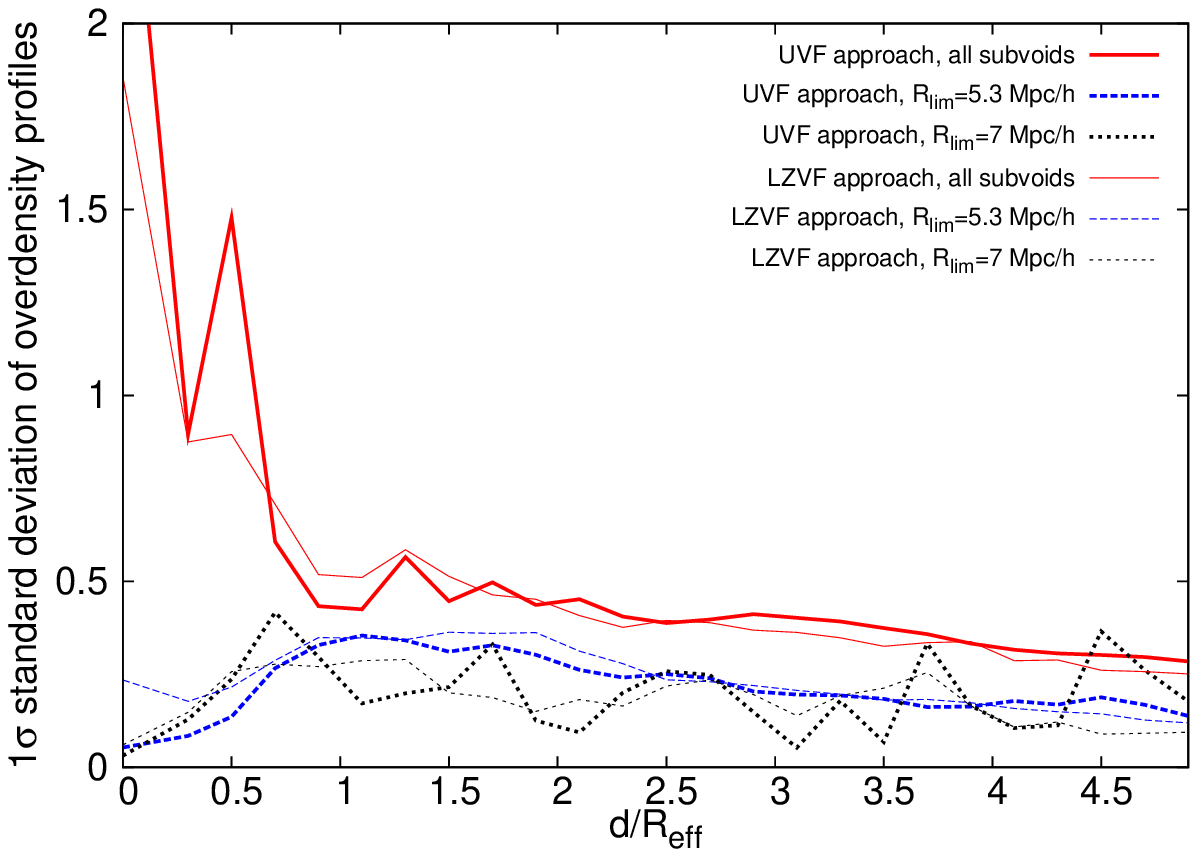,width=8.6cm}\\
\end{tabular}
\caption{Mean overdensity profiles ({\em left-hand panel}) and
  corresponding $1\sigma$ deviations ({\em right-hand panel}) of the
  voids detected by the two finders UVF and LZVF, as a function of the
  normalized radial distance, $d/R_{eff}$, and for three different
  values of $R_{lim}$ ({\bf M} centring), as indicated by the labels.}
\label{fig:density}
\end{figure*}

Fig.~\ref{fig:density} compares the mean overdensity profiles and
corresponding $1\sigma$ deviations of the subvoids, moderate
($R_{lim}=5.3~{\rm Mpc}/h$) and large voids ($R_{lim}=7~{\rm Mpc}/h$)
detected by our two finders, assuming the {\bf M} centring.  The
profiles appear quite smooth, flattening at around $2R_{eff}$. Most of
them are not compensated, differently from what is obtained using void
finders based on density measurements (see \S\ref{sub:zobov}). We find
also very few voids with positive overdensity in the central radial
bin, due to the cloud-in-void mode of the void hierarchy
\citep{2004MNRAS.350..517S}. The $1\sigma$ deviations of the
overdensity profiles decrease for larger values of $R_{lim}$,
especially at small radial distances. Overall, the mean void profiles
obtained with the two finders UVF and LZVF appear in good agreement.
Divergence, overdensity and shape parameters of the selected void
samples are reported in Table \ref{table:2}.

\begin{table*}
\caption{Main parameters of the selected void samples. $1st$ and $2nd$
  columns are the names of the void finders and the values of
  $R_{lim}$.  $3rd$ and $4th$ columns show the average amplitudes of
  overdensity and divergence profiles in the central part of voids
  within $0.2\cdot R_{eff}$ with $1\sigma$ standard deviation. The
  values in bracket show the ratio between the average values and
  their standard deviations (that is the significance of the signal).
  $5th$ and $6th$ columns show the radius range of the voids used for
  stacking and their numbers, respectively. $7th$ and $8th$ columns
  represent the average ellipticities $<e>$ of individual voids with
  $1\sigma$ standard deviation, and the ellipticities of stacked voids
  $e^{st}$, respectively.  Ellipticities are measured using the
  divergence field for UVF and LZVF voids, and using halo positions
  for {\small ZOBOV} voids.}
\label{table:2}
\begin{tabular}{cccccccc}
\hline
void finder &$R_{lim}$ & overdensity $\delta$ &  divergence $\Theta$ & radius range, Mpc & N & $<e_{1,2}>$, $<e_{1,3}>$ & $e_{1,2}^{st}$ and $e_{1,3}^{st}$ \\       
\hline   
UVF   & 0 &$-0.79 \pm2.71$  (0.3) &  $-1.53 \pm 0.80$  (1.9)& 4-7  & 212 & $0.82\pm0.11$, $0.58\pm0.18$ & $0.99\pm0.01$, $0.97\pm0.01$ \\
UVF & 5.3 &$-0.99 \pm0.05$  (20) &  $-2.47 \pm 0.57$  (4.3) & 7-10 & 33 & $0.79\pm0.12$, $0.52\pm0.21$ & $0.98\pm0.04$, $0.97\pm0.05$ \\ 
UVF   & 7 &$-0.99 \pm0.03$  (33) &  $-2.57 \pm 0.48$  (5.4) & $>$7 &7 & $0.70\pm0.16$, $0.47\pm0.09$ & $0.63\pm0.24$, $0.46\pm0.21$ \\ 
\hline
LZVF  & 0  &$-0.56\pm1.86$  (3.3)  & -1.68 $\pm$ 0.87 (1.7)  & 4-7 & 222 & $0.84\pm0.11$, $0.64\pm0.19$ & $0.99\pm0.01$, $0.98\pm0.01$ \\ 
LZVF  &5.3 &$-0.90\pm0.23$  (3.9)  & -2.78 $\pm$ 0.66 (4.2)  & 7-10 &50 & $0.81\pm0.11$, $0.61\pm0.17$ &$0.97\pm0.03$, $0.96\pm0.03$  \\ 
LZVF  & 7  &$-0.98\pm0.06$  (16.3)  & -3.29 $\pm$ 0.61 (5.4) & $>$7 & 15 &$0.79\pm0.10$, $0.47\pm0.18$ & $0.89\pm0.09$, $0.83\pm0.11$ \\ 
\hline
{\small ZOBOV} & & -0.87 $\pm$ 0.33 (2.6) & -0.37 $\pm$ 1.00 (0.4) & 7-10& 34 & $0.76\pm0.10$, $0.59\pm0.11$ & $0.94\pm0.04$, $0.89\pm0.04$\\

\hline
\end{tabular}

\end{table*}

%%%%%%%%%%%%%%%%%%%%%%%%%%%%%%%%%%%%%%%%%%%%%%%%%%%%%%%%%%%%%%%%%%%%%%%%%%%%%%%%%%%

\subsection{Void ellipticities}

To further investigate the characteristics of the selected samples, we
estimate the ellipticity of stacked voids.  We consider three
sub-samples: subvoids with $4<r_{eff}[{\rm Mpc}/h]<7$, moderate voids
with $7<r_{eff}[{\rm Mpc}/h]<10$ and $R_{lim}=5.3~{\rm Mpc}/h$, and
large voids with $r_{eff}>7~{\rm Mpc}/h$ and $R_{lim}=7{~\rm Mpc}/h$.
To measure the ellipticity in the divergence field, we use the second
moments of the $\Theta$ distribution at the void position.  The
  components of the inertial tensor are defined in Lagrangian space
  with respect to the void centres:
\begin{equation}
I_{i,j}=\sum \chi_{i}\chi_{j}\Theta({\bf q}) \, ,
\label{eq:inertial}
\end{equation}      
where the smoothed divergence field $\Theta$ is defined over the nodes
of a cubic grid with 1 Mpc/h spacing.  The sum runs over all grid
points 
%inside the void and 
within a distance of $0.7\cdot R_{max}$ from the
void centre, where $R_{max}$ is the radius of the largest selected
void. The quantity $\chi_{i}$ represents the $i$-th Cartesian
component of the grid point in Lagrangian space with respect to the void center. The conservative
choice of considering points within $0.7\cdot R_{max}$ was proposed by
\citet{2014arXiv1404.5618S} to balance between the need to consider a
large number of grid points to maximise the signal-to-noise ratio and
that of avoiding fluctuations that are effectively outside the void
boundaries.
%where the sum is over void volume within sphere $0.7\cdot R_{max}$, indexes $i,j=1,2,3$,
% coordinates $\chi_{1}$, $\chi_{2}$ and $\chi_{3}$ are Cartesian in Lagrangian space. 
%Inertial tensor is calculated using gridded divergence field $\Theta(x,y,z)$, with grid step $1~$Mpc/h.
%$R_{max}$ is the maximum radius of the selected voids.
%\citet{2014arXiv1404.5618S} showed that considering the central part of void within $0.7\cdot R_{max}$ } provides a
%high signal-to-noise ratio, while minimising the fluctuations caused
%by regions outside the voids. 
We define the ellipticity, $e$, as the
axial ratio of the eigenvalues $\Lambda_{i}$ of the matrix $I_{i,j}$
given by Eq.~\ref{eq:inertial}:
\begin{equation}
e_{1,2} = \sqrt{\Lambda_{2}/\Lambda_{1}},~e_{1,3} =
\sqrt{\Lambda_{3}/\Lambda_{1}} \, ,
\end{equation}      
where $\Lambda_{1}\geq\Lambda_{2}\geq\Lambda_{3}$. The same definition
has already been used by \citet{ 1991MNRAS.249..662P,
  1995MNRAS.273...30D, 1997ApJ...479..632S}.  The errors on the
stacked void ellipticities are computed using the jackknife technique
\citep{1982Efron}.  The average and stacked void ellipticities are
reported in Table \ref{table:2}. The average ellipticity and its
variance provide information on individual void shapes. Voids
detected by our two finders have similar ellipticities, almost
independent of $R_{lim}$.  The ratio between major and medium axes is
in the range $0.70-0.84$ with $rms\sim0.1$, and between major and
minor axes in $0.47-0.64$ with $rms\sim0.15$.  Voids look like
elongated tri-axial ellipsoids. Overall, the void shape appears
approximately similar, independently on which of our finders (UVF or LZVF) we use. Ellipticities of the
stacked voids are close to unity, as expected, because our sample is
not redshift-space distorted. Finally, we note that subvoids have on
average a more spherical shape than larger voids.

%%%%%%%%%%%%%%%%%%%%%%%%%%%%%%%%%%%%%%%%%%%%%%%%%%%%%%%%%%%%%%%%%%%%%%%%%%%%%%%%%%%

\subsection {Comparison with ZOBOV finder}
\label{sub:zobov}

We compare our results with those obtained with {\small ZOBOV}
\citep{2008MNRAS.386.2101N}, a publicly available and widely used void
finder algorithm that searches for depressions in the density
distribution of a set of points. Applying a Voronoi tessellation, the
algorithm associates a cell at each point, defined as the region that
is closer to that point than to any other point of the sample. {\small
  ZOBOV} then identifies the local density minima, searching for the
Voronoi-cells whose density is lower than that of all the other
adjacent cells.  The Voronoi-cells surrounding the local density
minima are eventually joined, adding cells with larger and larger
densities. Voids are identified as unions of these Voronoi-cells.

However, local density minima can be found also in overdense
regions. To exclude the latter, we consider only regions with a
overdensity minimum smaller than $-0.8$ \citep{2008MNRAS.386.2101N}.  To assess the statistical significance
for each void, {\small ZOBOV} provides also the {\em fakeness
  probability}, that is the probability that a void is simply
generated by Poisson fluctuations in the distribution of points. In
this analysis, we consider only the voids with a significance level
larger than $2\sigma$, as commonly assumed.

To assess {\small ZOBOV} void centres, we use the following
definition:
\begin{equation}
{\bf x_c}=\frac{1}{V_{void}}\sum_{i=1}^{N_V}x_i^{halo}\cdot V_i^{halo} \, ,
\end{equation}
where ${\bf x_c}$ is the baricentre, $x_i$ and $V_i^{halo}$ are the
position of the i-$th$ halo in the void and the volume of the associated Voronoi cell, respectively, and
$V_{void}$ is the volume provided by {\small ZOBOV}. The main
properties of the {\small ZOBOV} void catalogue are reported in Tables
\ref{table:1} and \ref{table:2}.  To compare our void finders with
{\small ZOBOV}, we consider subsamples selected with $R_{lim}=5.3
{\rm Mpc}/h$. This choice has no particular physical meaning. It
just ensures that the number and effective radii of the voids
selected with the three different finders are similar (see Table
\ref{table:1}), and thus the statistical errors are similar as
well.  The interesting issue of how to determine the optimal value
of $R_{lim}$ for a specific cosmological analysis is postponed to
a forthcoming paper.

\begin{figure*}
\begin{tabular}{c c}
\hspace*{-0.5cm}\epsfig{file=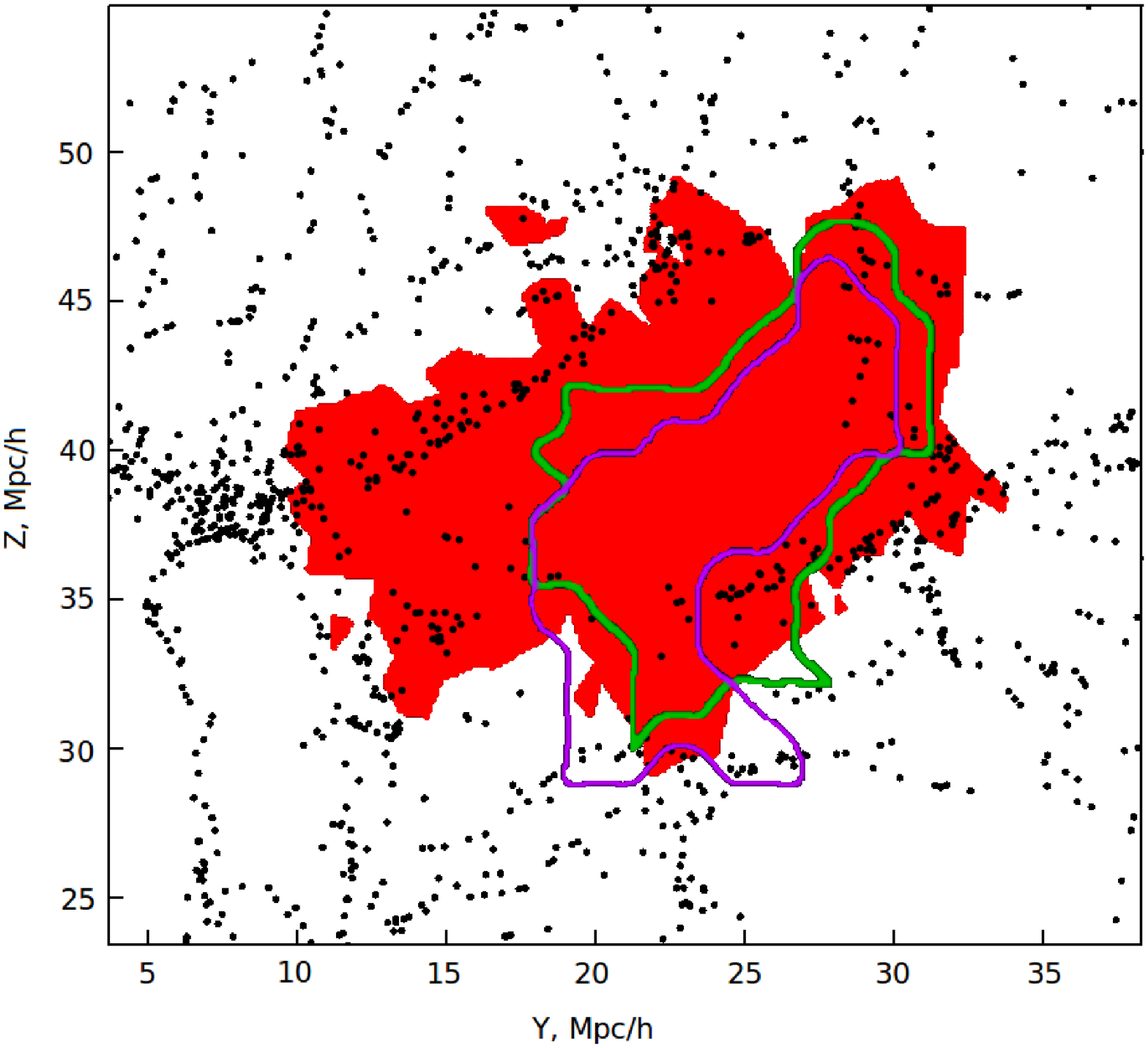,width=8.8cm}&\hspace*{0cm}\epsfig{file=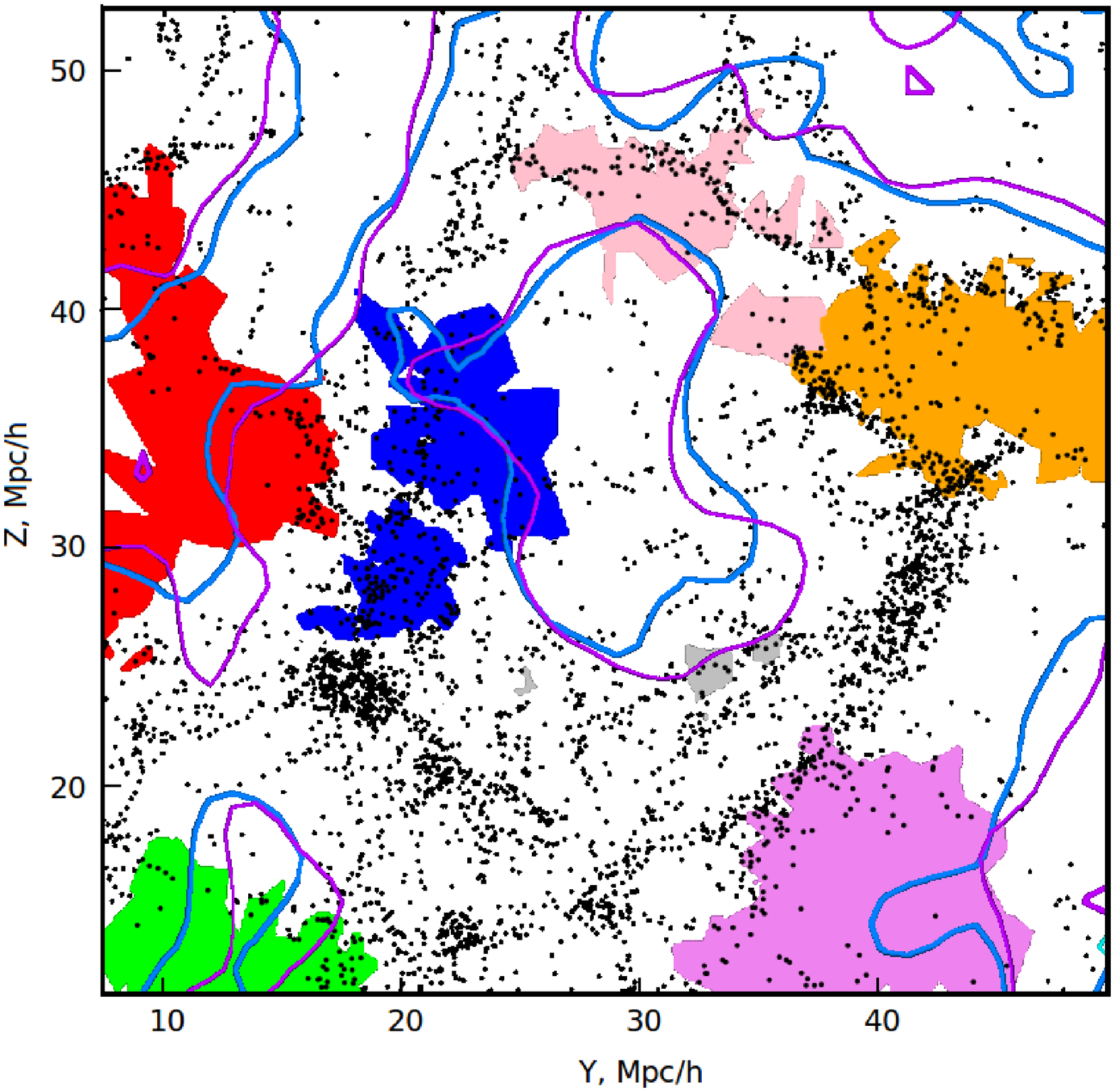,width=8.3cm}\\
\end{tabular}
\caption{Zoom of two regions of the slice shown in
  Fig.~\ref{fig:map}. Thick and thin lines show the shapes of voids
  selected by UVF and LZVF finders, respectively.
  Filled coloured areas show the voids found by {\small ZOBOV}. The
  underdense regions selected by UVF and LZVF appear always
  similar. On the other hand, voids selected by {\small ZOBOV} are in
  some cases similar to the ones detected by the other two finders
  ({\em left-hand panel}), while in other cases are very different
  ({\em right-hand panel}).}
\label{fig:zobov_map}
\end{figure*}

The black line in Fig.~\ref{fig:distr} shows the cumulative
distribution of the {\small ZOBOV} voids compared to our
findings. While the total numbers of voids selected by UVF, LZVF and
{\small ZOBOV} are similar, the {\small ZOBOV} voids appear
systematically larger than the others. The median radii of UVF, LZVF
(for $R_{lim}=5.3~{\rm Mpc}/h$) and {\small ZOBOV} are $7.0$, $7.3$
and $8.2$ ${\rm Mpc}/h$, respectively.  The filling factor of the
{\small ZOBOV} voids is $56.4\%$, larger than the ones of our finders,
$41\%$ for UVF and $47\%$ for LZVF.

A one-by-one comparison of the voids selected by the three finders is
rather complicated, if not impossible, since different methods can
select very different topological underdense structures. The two
zoomed regions displayed in Fig.~\ref{fig:zobov_map} show
representative examples of two extreme cases. On the left-hand panel
we zoom on a large void detected by all the three finders,
approximately at the same positions. On the contrary, the region shown
on the right-hand panel demonstrates that, in some cases, the
underdense regions selected by {\small ZOBOV} can be entirely
different than the ones detected by UVF and LZVF, while the latter are
always in close agreement with each other. This is due to the significant
differences in the methods. For example, the overdensity in the
central part of the large void selected by UVF and LZVF (and not by
{\small ZOBOV}) in the centre of the right-hand panel of
Fig.~\ref{fig:zobov_map} is $-0.70\pm0.07$, thus larger than the
overdensity threshold used in {\small ZOBOV}.

\begin{figure*}
\begin{tabular}{c c}
\epsfig{file=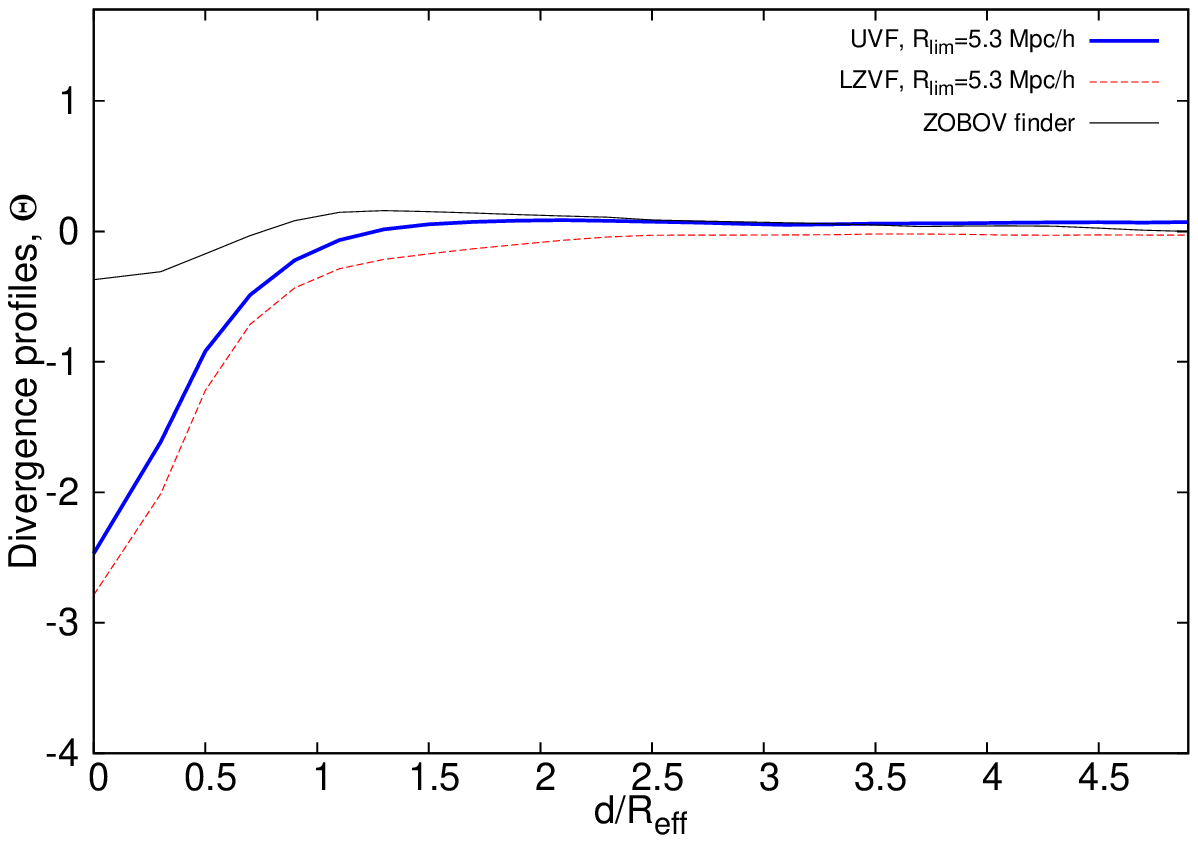,width=8.6cm}&\epsfig{file=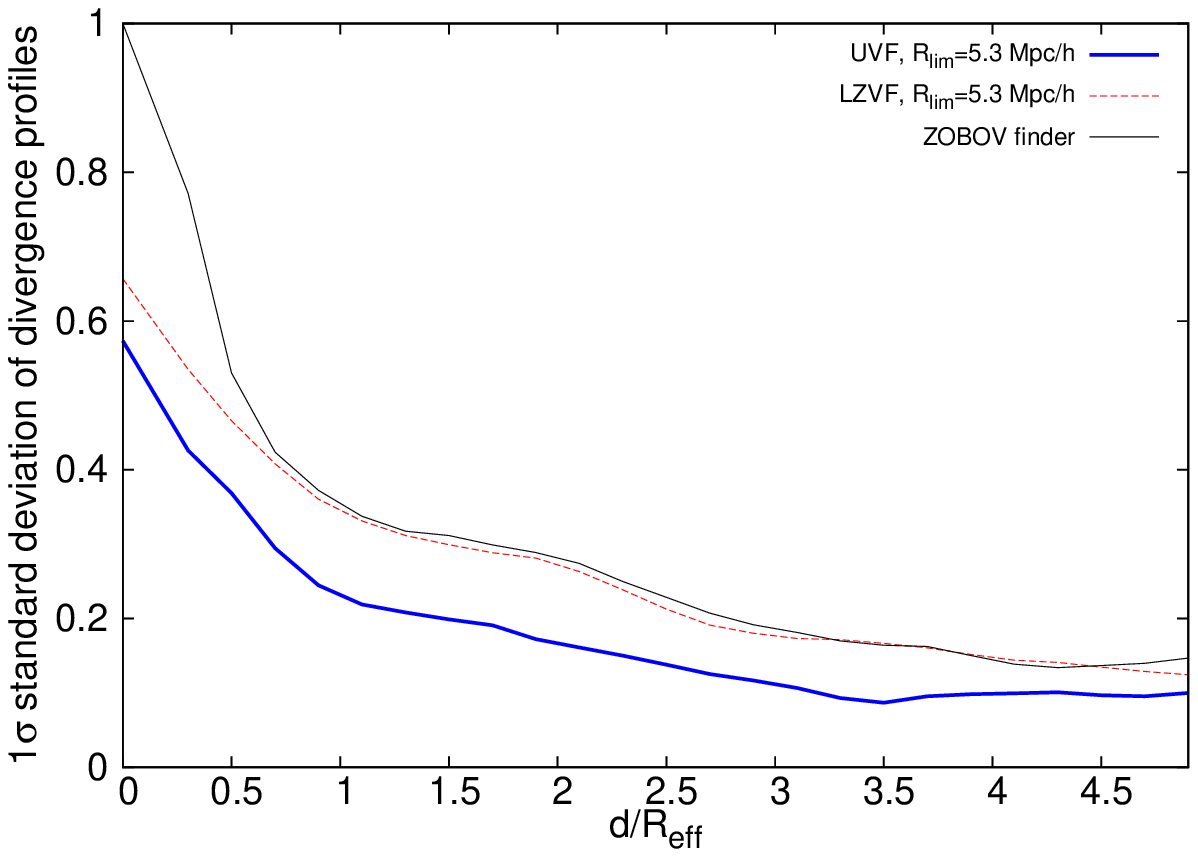,width=8.6cm}\\
\end{tabular}
\caption{Comparison between the mean divergence profiles ({\em
    left-hand panel}) and corresponding $1\sigma$ deviations ({\em
    right-hand panel}) of the voids detected by the two finders UVF
  and LZVF and the ones detected by {\small ZOBOV}, as a function of
  the normalized radial distance, $d/R_{eff}$.}
\label{fig:zoboz_divergence}
\end{figure*}

\begin{figure*}
\begin{tabular}{c c}
\epsfig{file=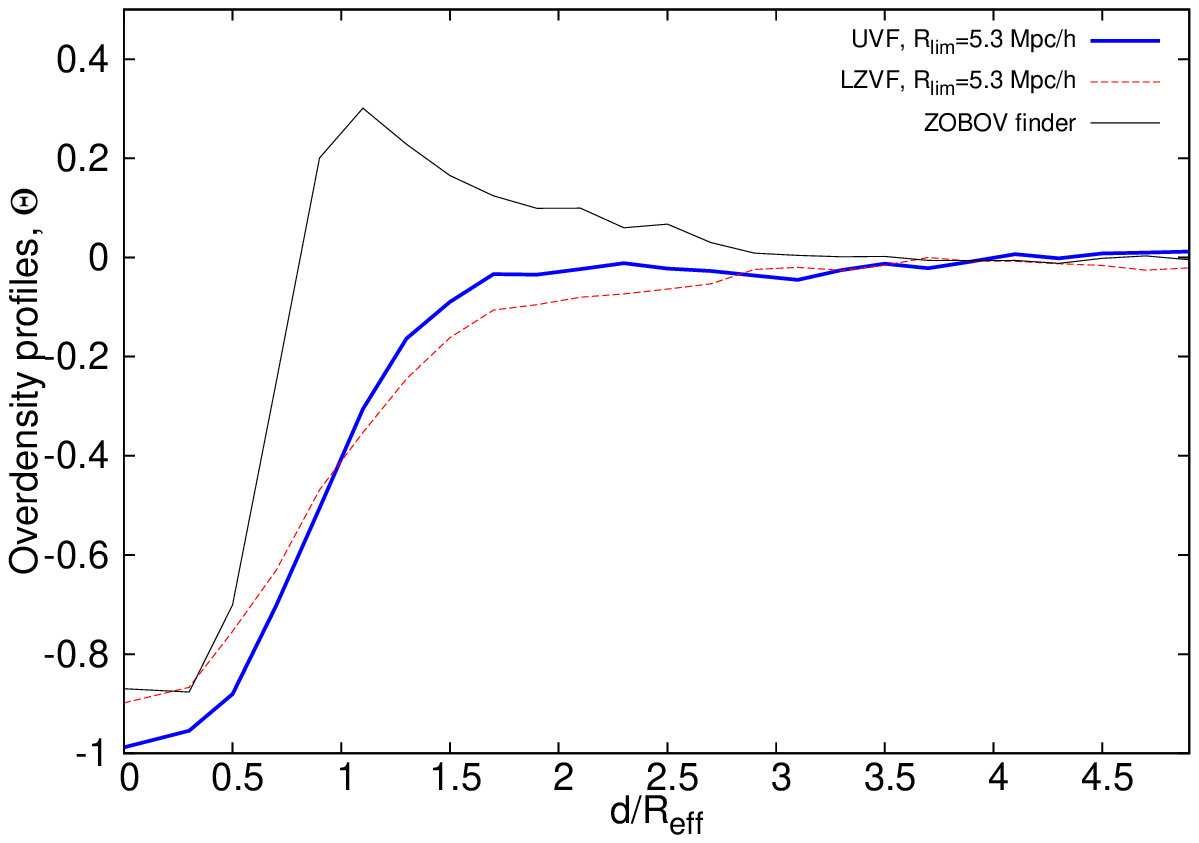,width=8.6cm}&\epsfig{file=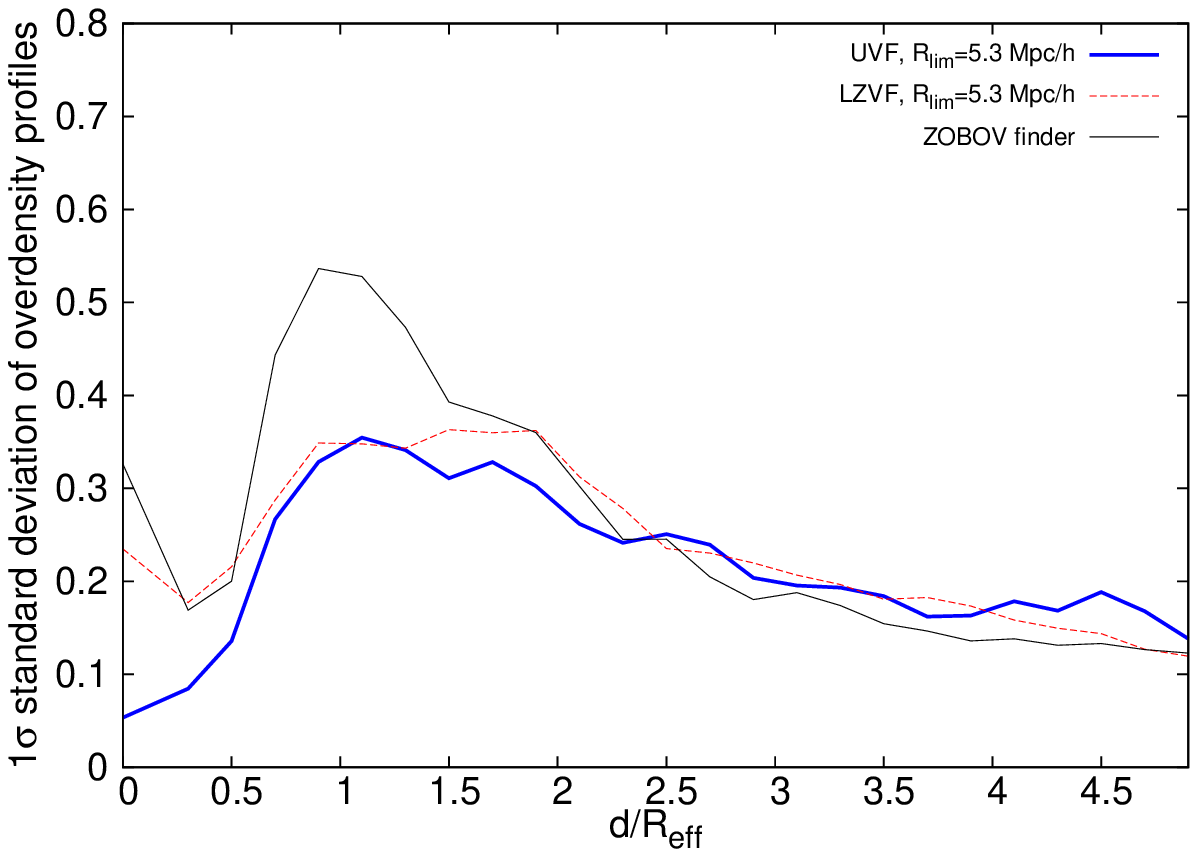,width=8.6cm}\\
\end{tabular}
\caption{Comparison between the mean overdensity profiles ({\em
    left-hand panel}) and corresponding $1\sigma$ deviations ({\em
    right-hand panel}) of the voids detected by the two finders UVF
  and LZVF and the ones detected by {\small ZOBOV}, as a function of
  the normalized radial distance, $d/R_{eff}$.}
\label{fig:zoboz_density}
\end{figure*}

Figs.~\ref{fig:zoboz_divergence} and \ref{fig:zoboz_density} compare
the mean divergence and overdensity profiles, and the corresponding
$1\sigma$ deviations, of the voids detected by the three finders UVF,
LZVF and {\small ZOBOV}. As it can be seen, the UVF and LZVF profiles are
steeper at small radial distances, and less scattered, especially the
UVF one, consistent with the universality of the void shapes
\citep{2014arXiv1407.1295N}. 
%Overall, {\small ZOBOV} overdensity
%profiles appear more compensated, {i.e. with a small overdensity} at
%$d/R_{eff}\simeq 1$. 
 The different profile shapes are caused by the
very different approaches of the analysed Eulerian and Lagrangian void
finders.

In Table \ref{table:2} we compare the significance of the signal at
the first radial bin, within $0.2\cdot R_{eff}$, for both overdensity
and divergence profiles. As it can be seen, the most prominent signal is
in the overdensity profiles for {\small ZOBOV}, and in the divergence
profiles for UVF and LZVF. Hovewer, the significance of the divergence signal obtained from both our finders is $60\%$ higher than for overdensity profiles
of {\small ZOBOV} voids. This may be explained by the fact that, for
the calculation of the divergence field and corresponding void
selection, we use randomised samples of haloes that are less biased by
shot-noise in the central, extremely underdense, void regions.

For the {\small ZOBOV} voids we calculate the inertial tensor using
the Eulerian coordinates of halos within $0.7\cdot R_{max}$  
%through Eq.~\ref{eq:inertial}, 
and assuming that all the haloes have equal masses, $m=1$:
\begin{equation}
I_{i,j}=\sum_{k} \chi_{i,k}\chi_{j,k} \, ,
\label{eq:inertial}
\end{equation}      
where the index $k$ corresponds to the index of the considered halo. 
UVF and LZVF have ellipticities closer to unity than {\small ZOBOV}
($e^{st}_{1,2}= 0.98$, $0.97$ and $0.94$, respectively), and the same
$rms\sim0.04$.  For the ratio of major and minor axes, we also find
systematically closer results to unity for UVF and LZVF than {\small
  ZOBOV} ($0.97$, $0.96$ and $0.89$, respectively).


%%%%%%%%%%%%%%%%%%%%%%%%%%%%%%%%%%%%%%%%%%%%%%%%%%%%%%%%%%%%%%%%%%%%%%%%%%%%%%%%%%%

\section{Summary and Conclusions}
\label{sec:conclustions}

Present and upcoming next generation galaxy redshift surveys, such as
SDSS \citep{2009ApJS..182..543A}, BigBOSS \citep{2009arXiv0904.0468S},
WFIRST \citep{2012arXiv1208.4012G}, HETDEX
\citep{2008ASPC..399..115H}, and Euclid \citep{2011arXiv1110.3193L},
will provide spectroscopic data with unprecedented large statistics,
allowing the use of underdense regions as effective cosmological
probes.  One of the most promising applications for cosmology is to
exploit void ellipticities \citep{2010MNRAS.403.1392L,
  2012ApJ...761..187S}.

The indentification of cosmic voids, however, is not trivial, due to their large
sizes and peculiar structures.  Generally, void finders are based on
density or geometric criteria, which can be affected by large shot
noise.  In this work we implemented two 
%new 
dynamical void finders, not affected by this weakness, and
characterised by a minimal number of free parameters and additional
assumptions.  The proposed void finders are based on sample
randomisation. One uses the Zel'dovich approximation to trace back in
time the orbits of matter tracers. The second uses the observed
two-point correlation function to relax the objects' spatial
distribution to homogeneity. The Lagrangian Zel'dovich Void
  Finder is similar to the {\small DIVA} procedure for what concerns
  the displacement field reconstruction \citep{2010MNRAS.403.1392L}.
  On the other hand, the Uncorrelating Void Finder adopts a new
  technique based on the measurement of the two-point correlation
  function of the sample. The two independent dynamical void finders
  considered in this work are thus complementary, and allowed us to
  strengthen our main results and conclusions.

 We defined voids as
sinks of the back-in-time streamlines of the mass tracers in
Lagrangian coordinates using the divergence of the displacement field.
With a watershed technique \citep{2007MNRAS.380..551P} we identified
hundreds of subvoids around the local minima of the divergence field.
These subvoids are parents of larger voids, which we constructed
using criteria of minimal effective void radius.  Moreover, we
considered four different approaches for the definition of void
centres, a crucial issue for a proper void stacking.  We found that
the most convenient option is the stacking of voids centred on their
minimum divergence. To test our finders we used a halo
catalogue from the {\small CoDECS} simulations.

We compared our results with those obtained with the publicly available Eulerian class
{\small ZOBOV} finder, using the same halo catalogue at $z=0$.
We found that the overdensity profiles of both UVF and LZVF are more
self-similar than the {\small ZOBOV} ones, thus their stacking should
provide more accurate cosmological probes.  The significance of the
signal in the central part of the voids ($<0.2R_{eff}$) is $60\%$
higher for the divergence profiles of our voids than for overdensity
profiles of {\small ZOBOV} voids, when using Eulerian halo
positions. We measured the ellipticities of both individual voids
and stacked ones.  We found a good agreement between individual void
shapes for all the three finders with average axis ratios
$<e_{1,2}>\sim0.8\pm0.1$, $<e_{1,3}>\sim0.55\pm0.15$. The void shape appears
approximately universal, independent of the fact that voids are
detected using overdensities or divergences.
  Stacked voids
from both our finders have much more spherical shape than {\small
  ZOBOV} ones.  The voids' properties of both our finders are in good
agreement with each other.  The two proposed void finders can thus
complement the existing set of finders and contribute to improve the
accuracy of cosmological probes.

The forthcoming Euclid survey will be very promising for the
application of our dynamical void finders.  The expected spectroscopic
survey will cover 15000 sq. deg. and will contain around $5\cdot
10^{7}$ galactic redshifts, mainly in the range $0.7<z<2.1$, up to visual
magnitude $m_{H}=24$ \citep{2011arXiv1110.3193L}. At such redshift range the mean
spacing will rise just from $4$ to $15~{\rm Mpc}/h$
\citep{2010PhRvD..82b3002B}.  This makes the Euclid spectroscopic
survey an ideal laboratory for the studying of voids' evolution.  The
volume of the Euclid spectroscopic survey will be larger than that of
SDSS by a factor of $500$. So we expect $300$ times more galaxies and
few hundreds times more voids in comparison to SDSS.

In the work by \cite{2010PhRvD..82b3002B} it has been shown that such
a high statistics of void ellipticities, combined with other
traditional methods, is expected to improve the DE task force figure of merit (FoM)
on the DE parameters by orders of hundreds. The simulation of
\cite{2012ApJ...754..109L} showed the effectiveness of the Euclid
stacked voids for the AP test and further DE probes.  The FoM of
stacked voids from the Euclid survey may double all the other DE probes
derived from the Euclid data alone (combined with Planck priors).
Moreover, the Euclid voids could in principle improve the outcomes
from baryon acoustic oscillations by an order of
magnitude. The cosmological constraints that can be obtained with the
CMB gravitational lensing on cosmic voids for a Euclid-like survey
have been investigated by \cite{2014arXiv1409.3364C}. The authors
found that the latter could be comparable to the constraints from
Planck data alone.

In a forthcoming paper we plan to further investigate the
characteristics of the cosmic voids detected by the proposed finders,
in particular the redshift evolution of the void number density, size
distribution and ellipticities, and their dependence on the
cosmological model.  We will also investigate the effect of
redshift-space distortions in the divergence field.

%%%%%%%%%%%%%%%%%%%%%%%%%%%%%%%%%%%%%%%%%%%%%%%%%%%%%%%%%%%%%%%%%%%%%%%%%%%%%%%%%%%

\section*{Acknowledgments}
We acknowledge the grants ASI n.I/023/12/0 "Attivit\`{a} relative alla
fase B2/C per la missione Euclid" and MIUR PRIN 2010-2011 "The dark
Universe and the cosmic evolution of baryons: from current surveys to
Euclid". MB is supported by the  Marie Curie Intra European Fellowship
``SIDUN"  within the 7th Framework  Programme of the European Commission. 
The numerical simulations presented in this work have been performed 
and analysed on the Hydra cluster at the RZG supercomputing centre in Garching, and on the BladeRunner cluster
at the Bologna University, partly funded by the Marie Curie Intra European Fellowship
``SIDUN".  We acknowledge financial support from PRIN INAF 2012 ”The Universe in the box: multiscale simulations of cosmic structure”.

%\clearpage
%\newpage

\end{document}